\documentclass[aps,pre,10pt,showpacs,floatfix,onecolumn,nofootinbib,a4paper]{revtex4-2}

\usepackage{bm,amsmath,empheq,amssymb,graphicx,stmaryrd,float,subcaption,upgreek,sidecap,dsfont,mathtools,gensymb,mathrsfs}
\usepackage[abs]{overpic}
\usepackage{hyperref,bookmark}
\usepackage{url}
\usepackage{comment,footnote}
\usepackage[format=plain,labelfont={bf,small},textfont=small,justification=raggedright,singlelinecheck=false]{caption}

\newcommand{\nigh}[1]{{\color{black}#1}}

\usepackage{xcolor}

\newcommand{\br}{{\bm r}}

\newcommand{\bx}{{\bm x}}
\newcommand{\be}{{\bm e}}
\newcommand{\by}{{\bm y}}
\newcommand{\bd}{{\bm d}}
\newcommand{\bu}{{\bm u}}
\newcommand{\bomega}{{\bm\omega}}
\newcommand{\bn}{{\bm n}}

\newcommand{\bI}{\mathbf{I}}
\newcommand{\bQ}{\mathbf{Q}}

\newcommand{\bF}{{\bf F}}
\newcommand{\bL}{{\bf L}}

\newcommand{\bq}{{\bm q}}

\newcommand{\bC}{{\bf C}}
\newcommand{\ba}{{\bm a}}
\newcommand{\bb}{{\bm b}}
\newcommand{\bnu}{{\bm \nu}}

\newcommand{\ave}[1]{{\left\langle #1 \right\rangle}}
\newcommand{\trans}{^\mathsf{T}}

\newcommand{\n}{\bm{n}}

\newcommand{\dv}{\bm{d}}

\newcommand{\e}{\bm{e}}
\newcommand{\framee}{\{\e_1,\e_2,\e_3\}}
\newcommand{\framed}{\{\dv_1,\dv_2,\dv_3\}}
\newcommand{\x}{\bm{x}}
\newcommand{\zero}{\bm{0}}
\newcommand{\y}{\bm{y}}
\newcommand{\normal}{\bm{\nu}}

\newcommand{\Binormal}{\bm{B}}
\newcommand{\Normal}{\bm{N}}
\newcommand{\Tangent}{\bm{T}}
\newcommand{\bR}{\bm{R}}
\newcommand{\step}{\mathbf{L}}
\newcommand{\Q}{\mathbf{Q}}
\newcommand{\surface}{\mathscr{S}}
\newcommand{\free}{\mathscr{F}}
\newcommand{\nablas}{\nabla\!_{\mathrm{s}}}
\newcommand{\f}{\bm{f}}
\newcommand{\directrix}{\bm{r}}
\newcommand{\C}{\mathbf{C}}
\newcommand{\Cf}{\C_{\f}}
\newcommand{\nay}{(\nabla\y)}
\newcommand{\curvature}{(\nablas\normal)}

\newcommand{\ribbon}{\mathscr{S}_0}
\newcommand{\slab}{\mathsf{S}}

\newcommand{\euclid}{\mathscr{E}}

\DeclareMathOperator{\tr}{tr}
\usepackage[utf8]{inputenc}
\usepackage[T1]{fontenc}
\usepackage{nomencl}

\cleardoublepage           
\renewcommand{\nomname}{} 
\begin{document}

	\title{A ribbon model for nematic polymer networks}
	\author{Harmeet Singh}
	\email{harmeet.singh@epfl.ch}	
	\affiliation{Laboratory for Computation and Visualization in Mathematics and Mechanics, Institute of Mathematics, \'Ecole Polytechnique F\'ed\'erale de Lausanne, 1015 Lausanne, Switzerland}
	\author{Epifanio G. Virga}	
	\email{eg.virga@unipv.it}
	\affiliation{Department of Mathematics, University of Pavia, Italy}
	\date{\today}

\begin{abstract}
We present a theory of deformation of ribbons made of nematic polymer networks (NPNs). 
These materials exhibit properties of rubber and nematic liquid crystals, and can be activated by external stimuli of heat and light.
A two-dimensional energy for a sheet of such a material has already been derived  from the celebrated neo-classical energy of nematic elastomers in three space dimensions.
Here, we use a dimension reduction method to obtain the appropriate energy for a ribbon from the aforementioned sheet energy.
We also present an illustrative example of a rectangular NPN ribbon that undergoes in-plane serpentine deformations upon activation under an appropriate set of boundary conditions. 
\end{abstract}
\maketitle

\vspace{-10pt}
\section{Introduction}\label{sec:introduction}
Photoactivation of elastic materials promises to make an old dream come true: to convey energy from a distance so as to induce matter to do work with no direct, material contact with the energy source. In particular, photoactive nematic elastomers seem especially promising materials. These are rubber-like elastic solids whose constituting polymer chains incorporate photoactive molecules into nematic liquid crystal elastomers.

A typical example is the azobenzene molecule, which goes from a rod-like to a V-shaped conformation (a process called the \emph{trans-cis} isomerization) when absorbing a photon of the appropriate frequency.\footnote{Thermal relaxation then induces the reverse shape change, restoring the molecule's ground state.} Current wisdom has it that the change in shape induced by the (few) photoactive molecules present in the polymer chains (mostly constituted by nematogenic rod-like molecules unaffected by illumination) has the potential to alter the nematic orientational order reigning among the latter molecules, in a way comparable to what a change in temperature may achieve. Completely disordered molecules render the polymer network isotropic, a symmetry inherited by the macroscopic elastic response. Ordered molecules will instead grant an anisotropic response, which to first approximation is described by the \emph{neo-classical} theory presented in the book by Warner and Terentjev \cite{warner:liquid}  (the abundant literature that precedes it \cite{blandon:deformation,warner:soft,terentjev:orientation,verwey:soft,verwey:multistage,verwey:elastic} is also of interest for the theoretician). 

An interesting mechanical theory for photoactive nematic elastomers, which blends Warner and Terentjev's with the classical approach to the ordering nematic phase transition by Maier and Saupe \cite{maier:einfache}, has recently been proposed in  \cite{bai:photomechanical} (see also \cite{corbett:nonlinear,corbett:linear,corbett:polarization} for the relevant statistical mechanics antecedents). A vast body of reviews is available that also includes the specific topic of photoactivation of nematic elastomers \cite{mahimwalla:azobenzene,ube:photomobile,white:photomechanical,ula:liquid,pang:photodeformable,kuenstler:light}, to which \cite{warner:topographic} should also be added, which is broader in scope and perhaps more germane to a mechanician's taste. General continuum theories for nematic elastomers are also available in the literature, mostly in the 3D Euclidean setting \cite{anderson:continuum,zhang:continuum,mihai:nematic}. As for specific applications of programmable liquid crystal elastomers, we refer the reader to a recent special issue \cite{korley:introduction}.

Our development here will not be tied to a specific mechanism capable of inducing  an ordering change in the nematogenic polymer chains that constitute the material. We shall be content to say that there is a scalar  parameter $S_0$ that characterizes the order in the reference configuration and another scalar parameter, $S$, which characterizes the order in the current configuration, the difference between them being, however produced (either by light of by heat), the drive for the spontaneous deformation of the body.

The ordering of nematic molecules in both the reference and current configurations is further described by the director fields $\n_0$ and $\n$, which represent the average molecular orientation in the corresponding configurations.\footnote{We shall see in Sect.~\ref{sec:thin} and Appendix~\ref{app:step_tensor} how $S$ and $\n$ are related to the step-length tensor $\step$ and to the quadrupolar order tensor $\Q$, two related measures of molecular order for nematic elastomers.} 

We shall assume that the reference configuration is where the cross-linking takes place, so that both $S_0$ and $\n_0$ are known at the start. $S$ is induced by external stimuli, whose origin and nature will not play a specific role in our development.

The director $\n$ can be tied to the deformation of the body in several ways, including complete independence and complete enslaving. The nematic elastomers that we shall consider are of a special type; for them the cross-linking is so tight that the nematic director remains enslaved to the deformation.\footnote{What this constraint precisely entails will be shown shortly below, in Sect.~\ref{sec:thin}.} These material are called \emph{nematic polymer networks} (NPNs).\footnote{This name was proposed in \cite{white:programmable}. Others prefer to call them \emph{nematic glasses} (see, for example, \cite{modes:disclination}).} 

Perhaps the most interesting manifestations of the ability of NPNs to produce changes in shape capable of doing work are achieved when they take the form of thin sheets. We represent one such sheet as a slab of thickness $2h$ which in its 
reference configuration $\slab$ extends itself symmetrically on both sides of a planar surface $\ribbon$. The director $\n_0$ is \emph{blueprinted} \cite{modes:blueprinting} on $\ribbon$ (in its own plane) and extended uniformly across the thickness of $\slab$, with the same scalar order parameter $S_0$.

External stimuli can act on the molecular ordering so as to change $S_0$ into $S$, in a programmable way. The system is thus kicked out of equilibrium and a spontaneous deformation ensues, which makes the elastic free energy attain its minimum under the changed circumstances.

An elastic free-energy density (per unit volume), $f_e$, was put forward in \cite{blandon:deformation}. This energy is delivered by a ``trace formula'', which was derived by assuming an anisotropic Gaussian distribution for the polymer chains in the rubber matrix.\footnote{An interesting extension of the classical formula can be found in \cite{kutter:tube}, which builds on Edward's \emph{tube model} \cite{edwards:theory} for entangled rubber elasticity.} This formula features both the deformation $\f$ of the three-dimensional slab $\slab$ and the (tensorial) measures of order in both the reference and current configurations (see Sect.~\ref{sec:thin}). For a sufficiently thin slab $\slab$, however, one would like to reduce $f_e$ to a function of the mapping $\y$ that changes the flat reference mid-surface $\ribbon$ into the orientable, curved surface $\surface$ in the current configuration that can be regarded as the core of the deformed slab $\f(\slab)$, although it need not be its mid-surface.

For a NPN, for which $f_e$ ultimately depends only on $\f$, such a dimension reduction was performed in \cite{ozenda:blend} by extending a standard method of the theory of plates, known as the Kirchhoff-Love hypothesis \cite{ozenda:kirchhoff}. As expected, this method delivers a surface elastic free-energy density with two components, a \emph{stretching} component $f_s$ scaling like $h$, and a \emph{bending} component $f_b$ scaling like $h^3$; $f_s$ depends only on the two-dimensional stretching (or metric) tensor $\C:=\nay\trans\nay$, while $f_b$ also depends on the invariant measures of curvature of $\surface$ and the relative orientation of $\n$ in the frame of the principal directions of curvatures of $\surface$.  

Not only do $f_s$ and $f_b$ scale differently with $h$, they also bear a different meaning in regard to the embedding of $\surface$ in space. By Gauss' \emph{theorema egregium} \cite[p.\,139]{stoker:differential}, the Gaussian curvature $K$ of $\surface$ is fully determined by the metric tensor $\C$, thus deserving the name of \emph{intrinsic} curvature, whereas other measures of curvature of $\surface$, not determined by the metric, are called \emph{extrinsic} and are affected by how $\surface$ is embedded in space.

Now, $f_s$ depends only on the intrinsic curvature, whereas $f_b$ also depends on the extrinsic ones. Studying the equilibria of $\surface$ under the \nigh{\emph{blended}} effect of both energy components $f_s$ and $f_b$ has proven a formidable task; a number of approximations have been proposed \cite{pedrini:ridge,pedrini:ridge_JPA}, but no satisfactory unified, general treatment of the blended surface energy  has so far become available.

In this paper, we shall move yet a further step in the dimensional reduction cascade, by considering the \emph{ribbon} limit for $\ribbon$ and its activated, deformed companion $\surface$. We shall describe $\ribbon$ as generated by a centerline $\directrix_0$ and the imprinted director field $\n_0$, which in this limit reduces to a unit vector field defined on $\directrix_0$. As a consequence, $\surface$ too will be described by a curve $\directrix$ in space, enriched with a unit vector field $\n$, which is $\n_0$ entrained by the deformation $\y$ of $\ribbon$. \nigh{In our model, in accord with a large body of literature (see, for example, Sect.\,5.2 of \cite{audoly:elasticity}), a} NPN ribbon is a \emph{decorated} curve retaining the essential geometric ingredients to represent a two-dimensional body, with one side much smaller than the other \nigh{and retained only at the lowest order. This implies that the bend of the nematic director field vanishes both in the reference and deformed configurations. A different, more elaborate model of ribbon, contemplating non-vanishing nematic bend and accounting for higher-order terms in the ribbon's width, was proposed in \cite{grossman:elasticity} starting from a membrane Hamiltonian, which however appears to be divorced from the trace formula of three-dimensional elastomers, which is our main thrust here.}

What is characteristic of \nigh{our model} compared to the varieties studied in classical elasticity \nigh{\cite[Chap.\,5]{audoly:elasticity}} is the material nature of the directors $\n_0$ and $\n$: they are linked to the nematic order of molecules present in the polymer network. \nigh{Other studies on ribbons of activable elastomers populate the recent literature. We refer, in particular, to \cite{agostiniani:shape}, where a ribbon model is derived via $\Gamma$-convergence from a plate energy \cite{agostiniani:rigorous} phrased in the language of non-Euclidean elasticity \cite{efrati:elastic}. Besides the method used, the major difference between these studies and ours lies in the assumption about the imprinted director $\n_0$, which is uniform throughout the cross-section of the parent sheet in our setting, whereas it is not uniform in those others, being there a possible source for kinematic incompatibility.}

Our main purpose here is to derive the ribbon energy for a NPN from the plate energy obtained in \cite{ozenda:blend} with no specific assumption on the ribbon's geometry. This goal is achieved in Sect.~\ref{sec:energy}, after having set the necessary kinematic preliminaries in Sect.~\ref{sec:kinematics}. Section~\ref{sec:geometry} is devoted to a special geometric setting: we study a rectangular ribbon, for which we find the explicit equilibrium planar solutions for a class of boundary conditions compatible with activable \emph{serpentine} modes. We shall see how an activated serpentine shape of the ribbon is determined by the director field $\n_0$ imprinted in its reference configuration. Finally, in Sect.~\ref{sec:conclusion}, we collect our conclusions and comment on the implications of our work and its possible future extension. The paper is closed by three technical appendices, where details of our development are expounded for the ease of the demanding reader.

\section{Thin sheet energy}\label{sec:thin}
Following in part \cite{pedrini:ridge}, we recall in this section the plate-like theory for NPNs obtained in \cite{ozenda:blend} through the dimension reduction method illustrated in \cite{ozenda:kirchhoff}.\footnote{The reader will find in \cite{mihai:plate} an alternative plate theory for nematic elastomers.} This method was applied to the ``trace formula'' for the elastic free-energy density (per unit volume) that had been put forward for nematic elastomers  \cite{blandon:deformation,warner:theory,warner:elasticity,warner:nematic_elastomer} (see also \cite[Chapt.\,6]{warner:liquid} for a comprehensive account of this theory) as an extension to anisotropic solids of the classical Gaussian theory for rubber elasticity (racapitulated in the landmark book \cite{treloar:non-linear_third}). 

Two director fields feature in this theory, $\n_0$ and $\n$, the former defined in the reference configuration of the slab $\slab$ and the latter defined in the current configuration $\f(\slab)$, where $\f$ is a diffeomorphism of $\slab$ in three-dimensional Euclidean space $\euclid$. In each configuration, the corresponding director represents the average orientation of the nematogenic molecules appended to the rubber polymeric matrix. They are more properly defined through the tensorial measures of material anisotropy that describe the end-to-end Gaussian distribution of polymer strands. These are the \emph{step-length} tensors $\bL_0$ and $\bL$, in the reference and current configurations, respectively, which, following \cite{verwey:elastic} and \cite{nguyen:theory}, we write as 
\begin{equation}
	\label{eq:step_tensor}
	\bL_0=A_0(\bI+S_0\bn_0\otimes\bn_0)\quad\text{and}\quad\bL=A(\bI+S\bn\otimes\bn).
\end{equation}
Here $\bI$ is the identity (in three-dimensional space), $A_0$ and $A$ are  positive geometric parameters (representing the persistence length perpendicular to $\n_0$ and $\n$, respectively), $S_0$ and $S$ are nematic scalar order parameters (related to the Maier-Saupe scalar order parameter, as shown in Appendix~\ref{app:step_tensor}).

The \emph{neo-classical} theory of nematic elastomers expresses the elastic free-energy density $f_e$ (per unit volume in the reference configuration) as
\begin{equation}
	\label{eq:f_e}
	f_e=\frac12k\tr(\bF\trans\bL^{-1}\bF\bL_0),
\end{equation}
where $\bF:=\nabla\f$ is the deformation gradient and $k>0$ is an elastic modulus. This is quite broadly known as the \emph{trace formula}.

In nematic polymer networks, $\n$ is enslaved to $\bF$. In these materials, with which we are concerned in this paper, the director field $\n_0$ is \emph{blueprinted} in the elastic matrix \cite{modes:blueprinting} and conveyed by the deformation into $\n$, which is thus delivered by
\begin{equation}\label{eq:n}
	\n=\frac{\bF\n_0}{|\bF\n_0|}.
\end{equation}
In general, elastomers are \emph{incompressible}, and so $\bF$ must satisfy
\begin{equation}\label{eq:incompressibility_F}
	\det\bF=1.
\end{equation}
Both \eqref{eq:n} and \eqref{eq:incompressibility_F} will be enforced as constraints on all admissible deformations $\f$ of $\slab$.

With $\n_0$ (and $S_0$) imprinted in the reference configuration at the time of cross-linking and $\n$ enslaved to the deformation, the only residual freedom lies with $S$, which can be changed by either thermal or optical stimuli. 

It was shown in \cite{ozenda:blend} that by use of \eqref{eq:step_tensor} and \eqref{eq:n} $f_e$ can be given the following form
\begin{equation}\label{eq:F_definition}
	f_e=\frac12k\frac{A_0}{A}F(\Cf),
\end{equation}  
where $\Cf:=\bF\trans\bF$ is the right Cauchy-Green tensor associated with the deformation $\f$ and 
\begin{equation}\label{eq:bulk_energy_density}
	F(\Cf):=\tr\Cf+\frac{S_0}{S+1}\n_0\cdot\Cf\n_0-\frac{S}{S+1}\frac{\n_0\cdot\Cf^2\n_0}{\n_0\cdot\Cf\n_0}.
\end{equation}
It is not difficult to show (see, for example, \cite{pedrini:ridge}) that $F$ subject to \eqref{eq:incompressibility_F} is minimized by 
\begin{equation}
	\label{eq:Cf_representation}
	\Cf=\lambda_f^2\n_0\otimes\n_0+\frac{1}{\lambda_f}(\bI-\n_0\otimes\n_0),
\end{equation} 
where
\begin{equation}
\lambda_f=\sqrt[3]{\frac{S+1}{S_0+1}},
\end{equation} 
which shows how spontaneous deformations can be induced in these materials.
For example, by heating the sample above the cross-linking temperature, we reduce the nematic order of the chains, so that $S<S_0$. This in turn induces a spontaneous deformation so as to minimize the total elastic free energy: fibers along $\n_0$ are shortened, whereas those in the plane orthogonal to $\n_0$ are dilated. Clearly, the reverse behaviour is expected upon cooling. Thus, $S$ can be regarded as the \emph{activation parameter} of our theory, driven by external stimuli. For definiteness, we shall conventionally assume that both $S_0$ and $S$ range in the interval $(-1,1)$.\footnote{Although, as shown by \eqref{eq:identifications}, the upper limit could be much larger.}

Here we are interested in thin sheets and in the appropriate dimension reduction of $F(\Cf)$ to be attributed to the mid surface $\ribbon$ of the slab $\slab$ of thickness $2h$. Formally, $\ribbon$ is a flat region in the $(x_3,x_1)$ plane of a (movable) Cartesian frame $\framee$ with $\e_2$ fixed in space, and $\slab$ is the set in $\euclid$ defined as $\slab:=\{(\x,x_2)\in \ribbon\times[-h,h] \}$. The mapping $\y:\ribbon\to\euclid$ describes the deformation of $\ribbon$ into the orientable surface $\surface=\y(\ribbon)$ in the deformed slab $\f(\slab)$; we shall assume that $\y$ is of class $C^2$ and that $\n_0$ is a two-dimensional field imprinted on $\ribbon$, so that $\n_0\cdot\e_2\equiv0$ (see Fig.~\ref{fig:general_schematics}).\footnote{In $\slab$, $\n_0$ is extended uniformly away from $\ribbon$, so as to be independent of the $x_2$ coordinate.}

Applying \eqref{eq:n} to the present setting, we obtain that
\begin{equation}
	\label{eq:n_y_setting}
	\n=\frac{(\nabla\y)\n_0}{|(\nabla\y)\n_0|}.
\end{equation}

In \cite{ozenda:blend}, we extended the classical Kirchhoff-Love hypothesis \cite{ozenda:kirchhoff} to obtain a dimension reduction of $F(\Cf)$ in \eqref{eq:bulk_energy_density}, that is, a method that convert $f_e$ in \eqref{eq:F_definition} into a surface energy-density $\tilde{f}_e$ (to be integrated over $\ribbon$). As standard in the theory of plates, such a surface energy is delivered by a polynomial in odd powers of $h$, conventionally truncated so as to retain the first two relevant ones, the first and the third power. The former is the \emph{stretching} energy $f_s$, accounting for the work done to alter distances and angles in $\ribbon$, while the latter is the \emph{bending} energy $f_b$, accounting for the work done to fold $\ribbon$. Thus, dropping the scaling constant $\frac12k\frac{A_0}{A}$, which has the physical dimensions of an energy per unit volume, we can write
\begin{equation}
	\label{eq:energy_splitting}
	\tilde{f}_e=f_s+f_b+O(h^5),
\end{equation}
where (\nigh{up to} an inessential additive constant)
\begin{subequations}
	\begin{eqnarray}
		f_s&=&\frac{2h}{\nigh{S}+1}\left(\tr\C+S_0\n_0\cdot\C\n_0+\frac{S}{\n_0\cdot\C\n_0}\right),\label{eq:f_s}\\
		f_b&=&\frac{2h^3}{3}\left\{2(8H^2-K)+\frac{1}{S+1}\left[\left(\frac{3S}{a_0^2}-a_0^2S_0-\tr\C\right)K-\frac{4S}{a_0^2}(2H-\kappa_n)\kappa_n\right] \right\}.\label{eq:f_b}
	\end{eqnarray}
\end{subequations}
Here  $\C:=\nay\trans\nay$ is the two-dimensional stretching (or metric) tensor, $a_0^2 := \n_0\cdot\C\n_0$, $H$ and $K$ are the mean and Gaussian curvatures of $\surface$, defined as
\begin{equation}
	\label{eq:H_K_definitions}
	H:=\frac12\tr\curvature\quad\text{and}\quad K:=\det\curvature
\end{equation}
in terms of the (two-dimensional) curvature tensor $\nablas\normal$, where $\normal$ is a unit normal field to $\surface$, and
\begin{equation}
	\label{eq:11}
	\kappa_n:=\n\cdot\curvature\n.
\end{equation}
The (scaled) total elastic free energy then reduces to the functional
\begin{equation}
	\label{eq:free_energy_functional}
	\free[\y]:=\int_{\ribbon}(f_s+f_b)dA,
\end{equation}
where $A$ is the area measure.

\section{Ribbon kinematics}\label{sec:kinematics}

We first establish the kinematics of the deformation of a planar ribbon $\ribbon$.
\begin{figure}[h]
	\centering
	\includegraphics[width=\textwidth]{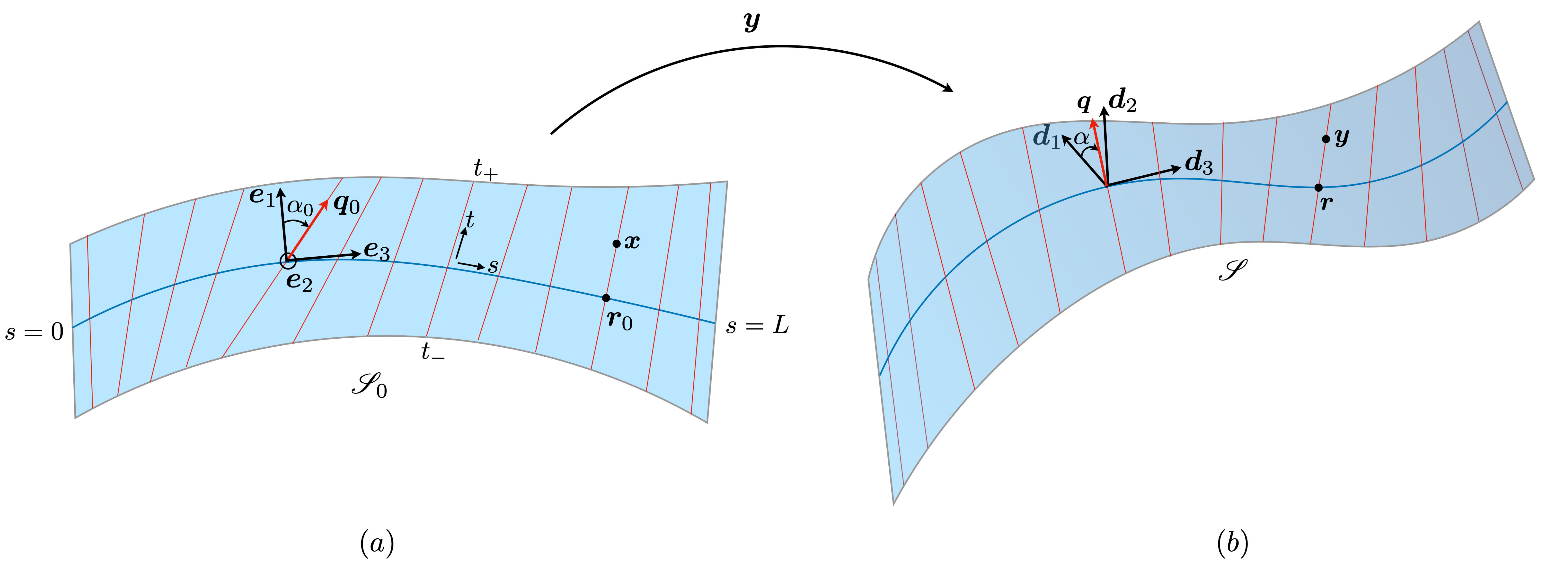}
	\caption{(a) The reference configuration $\ribbon$ of a planar ribbon, ruled by the directrix $\br_0(s)$ and the generators in the direction of the unit vector $\bq_0(s)$. An orthonormal set of directors $\{\be_1(s),\be_2,\be_3(s)\}$ are adapted to the centreline such that $\be_3(s)$ is tangential to $\br_0(s)$, and $\be_2$ is pointing out of the plane of the paper. The angle between $\bq_0$ and $\be_1$ is denoted by $\alpha_0$. (b) A current, non-planar, configuration $\mathscr{S}$ of the ribbon with the directrix denoted by $\br(s)$, and the direction of the generators given by $\bq(s)$. An orthonormal triad $\{\bd_1(s),\bd_2(s),\bd_3(s)\}$ is attached to the centreline with $\bd_3(s)$ oriented along the tangent vector, and $\bd_1(s)$  in the tangent plane at $s$. The angle between $\bd_1(s)$ and $\bq_0(s)$ is shown by $\alpha$.}
	\label{fig:general_schematics}
\end{figure}
We choose on $\ribbon$ a material line $\br_0(s)$, where $s$ is its arc-length coordinate.
We call this line the directrix in consonance with the mathematical terminology of ruled surfaces.
The centreline is endowed with an ordered orthonormal frame of directors $\{\be_1(s),\be_2,\be_3(s)\}$ oriented such that, $\be_3(s) = \partial_s\br_0(s)$, $\be_2$ is a constant vector pointing out of the plane of the surface, and $\be_1(s):=\be_2\times\be_3(s)$. 
Due to the orthonormality of the frame, we can associate with it a Darboux vector $\bomega(s)$ such that it satisfies $\partial_s\be_i = \bomega\times\be_i$, $i=1,2,3$. Since $\br_0(s)$ is a planar curve, $\bomega$ can be fully described by just one component in the director frame, i.e. $\bomega = \omega_2\be_2$.

The planarity of the reference configuration allows us to describe it using the following parametrisation of a ruled surface,\footnote{Which is standard in the general theory of ribbons, as recounted for example in \cite{dias:wunderlich} (see also the discussions in \cite{chen:moebius,chen:issues,heijden:comment,chen:reply}, not necessarily in unison).}
\begin{align}
\bx(s,t) = \br_0(s) + t \bq_0(s)\, ,\label{eq:reference_representation}
\end{align}
 where $\bq_0(s)$ is a unit vector given by,
\begin{align}
\bq_0(s) = \cos\alpha_0(s)\be_1 + \sin\alpha_0(s)\be_3 \quad\text{with}\quad -\frac{\pi}{2}\le\alpha_0\le\frac{\pi}{2}\, ,\label{eq:q0}
\end{align}
and $t$ is a coordinate along $\bq_0$.
We also define for later use,
\begin{align}
\bq_0^\perp:=\be_2\times\bq_0 = -\cos\alpha_0\be_3 + \sin\alpha_0\be_1\, .\label{eq:q0_perp}
\end{align}
Hereafter, $\bq_0$ will be identified with the imprinted nematic director $\n_0$, and so it acquires a material significance,
\begin{equation}
	\label{eq:q_0=n_0}
	\n_0=\bq_0.
\end{equation}
A planar strip on $\ribbon$ is represented in the $(s,t)$ coordinates as the set $\{(s,t):0\le s \le L, t_{-}(s)\le t\le t_{+}(s)\}$, where the functions $t_{\pm}(s)$ are such that $\bx(s,t_\pm)$ corresponds to the edges of the strip (see Fig. \ref{fig:general_schematics}). 

Upon deformation, the centreline $\br_0(s)\in \ribbon$ is convected into a curve $\br(s)\in \mathscr{S}$. 
Similar to $\br_0(s)$, we associate an ordered orthonormal director frame $\{\bd_1(s),\bd_2(s),\bd_3(s)\}$ with $\br(s)$ such that
\begin{align}
\bd_3(s) := \frac{\partial_s\br}{v_3}\, ,\label{eq:stretch_relation}
\end{align}
with $v_3(s):=|\partial_s\br|$ defined as the stretch of the centerline. 
The orthonormality of the directors $\bd_i$, $i=1,2,3$, allows us to associate with it a  Darboux vector $\bu(s)$ such that $\partial_s\bd_i=\bu\times\bd_i$. 
The director components of $\bu(s)=u_1\bd_1 + u_2\bd_2 + u_3\bd_3$ represent the bending strains about the corresponding directors.

The deformation of the reference configuration $\ribbon$ into the current configuration $\mathscr{S}$ is represented by the following map,
\begin{align}
\by(s,t) = \br(s) + \psi(s,t)\bq(s)\, ,\label{eq:y_representation}
\end{align}
where  $\bq(s)$ is a unit vector given by,
\begin{align}
\bq(s) = \cos\alpha(s)\bd_1(s) + \sin\alpha(s)\bd_3(s)\quad\text{with}\quad -\frac{\pi}{2}\le\alpha\le \frac{\pi}{2}\, .\label{eq:q}
\end{align}
\nigh{Equation \eqref{eq:y_representation} has an antecedent in equation (1) of \cite{chen:representation_2015}, whose style and notation are often echoed here.}

To obtain the deformation gradient induced by the mapping \eqref{eq:y_representation}, we consider a curve in $\ribbon$ represented by $(s(\xi),t(\xi))$, parametrised by some parameter $\xi$.
On differentiating \eqref{eq:reference_representation} and \eqref{eq:y_representation} with respect to $\xi$, it follows that the curve and its image in the current configuration $\mathscr{S}$ satisfy,
\begin{subequations}\label{eq:x_y_dots}
\begin{align}
	\dot{\bx} &= (\partial_s\br_0 + t\partial_s\bq_0)\dot{s} + \bq_0\dot{t}\, ,\label{x_dot}\\
	\dot{\by} &= (\partial_s\br + \partial_s\psi\bq + \psi\partial_s\bq)\dot{s} + \partial_t\psi\bq\dot{t}\, ,\label{eq:y_dot}
\end{align}
\end{subequations}
where the superimposed dot denotes differentiation with respect to $\xi$. 
The deformation gradient $\nabla\by$ can be represented as,
\begin{align}
\nabla\by = \ba\otimes\bq_0 + \bb \otimes\bq_0^\perp\, ,\label{eq:deformation_gradient}
\end{align}
where vectors $\ba$ and $\bb$ are respectively the images of $\bq_0$ and $\bq_0^\perp$ in the current configuration. 
Requiring that the identity $\dot{\by} = \left(\nabla\by\right)\dot{\bx}$ be valid for all $(\dot{s},\dot{t})$, we obtain using \eqref{eq:x_y_dots} and \eqref{eq:deformation_gradient},
\begin{subequations}
	\begin{align}
	\ba &= \partial_t\psi\bq\, ,\label{eq:a}\\
	\bb &= \frac{\partial_s\br + (\partial_s\psi - \partial_t\psi\partial_s\br_0\cdot\bq_0)\bq + \psi\partial_s\bq}{\bq_0^\perp\cdot\partial_s\br_0 + t\bq_0^\perp\cdot\partial_s\bq_0}\, .\label{eq:b_temp}
	\end{align}
\end{subequations}
Using $\partial_s\br_0 = \be_3$, and equations \eqref{eq:q0}, \eqref{eq:q0_perp}, and \eqref{eq:stretch_relation}, we give \eqref{eq:b_temp} the following simpler form,
\begin{align}
\bb=-\frac{1}{\cos\alpha_0+t(\partial_s\alpha_0-\omega_2)}[v_3\bd_3+(\partial_s\psi-\partial_t\psi\sin\alpha_0)\bq+\psi\partial_s\bq].\label{eq:b}
\end{align}

With the kinematics of the deformation established, we next enforce the constraint of inextensibility on the material of the ribbon, and compute its consequences.

\subsection{\nigh{Incompressibility} constraint}\label{sec:incompressiblity_constraint}
Nematic elastomers are subject to the constraint \eqref{eq:incompressibility_F}.
Using \eqref{eq:deformation_gradient}, \eqref{eq:a}, and \eqref{eq:b}, this constraint is also expressed as
\begin{align}
|\ba\times\bb|^2 = \frac{(\partial_t\psi)^2}{\left[\cos\alpha_0+t(\partial_s\alpha_0-\omega_2)\right]^2}|\bq\times(v_3\bd_3+\psi\partial_s\bq)|^2=1\, .\label{eq:inextensibility}
\end{align}
To solve \eqref{eq:inextensibility} we first compute the following useful relations,
\begin{align}
	\bq\times\bd_3 = -\cos\alpha\bd_2\, ,\qquad 
	\bq\times\partial_s\bq=(u_1\sin\alpha - u_3\cos\alpha)\bq^\perp + (u_2 - \partial_s\alpha)\bd_2\, ,\label{eq:computations}
\end{align}
where $\bq^\perp:=\bd_2\times\bq$. 
Making use of \eqref{eq:computations}, we see that the most general solution of \eqref{eq:inextensibility} is the following,
\begin{subequations}
	\label{eq:general_solution}
	\begin{align}
	\psi&=a(s) t\quad\text{with}\quad a(s)\geqq\frac{\cos\alpha_0}{v_3}>0,\label{eq:general_solution_1}\\
	\alpha&=\arccos\left(\frac{\cos\alpha_0}{v_3 a}\right),\label{eq:general_solution_2}\\
	u_2&=\partial_s\alpha-\frac{\partial_s\alpha_0-\omega_2}{a^2},\label{eq:general_solution_3}\\
	u_3&=u_1\tan\alpha.\label{eq:general_solution_4}
	\end{align}
\end{subequations}
Details of the computations leading to \eqref{eq:general_solution} are outlined in Appendix \ref{app:general_solution}.  
One consequence of \eqref{eq:general_solution} is the following useful relation,
\begin{align}\label{eq:consequence}
\partial_s\bq = -\frac{\partial_s\alpha_0-\omega_2 }{a^2} \bq^\perp\, ,
\end{align}
which we shall repeatedly invoke later.
Correspondingly, we see that relations \eqref{eq:general_solution}, upon using \eqref{eq:consequence}, simplify the representation for $\ba$ and $\bb$ from \eqref{eq:a} and \eqref{eq:b} to the following,
\begin{subequations}
	\label{eq:a_b_solutions}
	\begin{align}
	\ba&=a\bq,\label{eq:a_b_solutions_a}\\
	\bb&=-\left[\frac{v_3\sin\alpha-a\sin\alpha_0+t \partial_s a }{\cos\alpha_0+t(\partial_s\alpha_0-\omega_2)}\right]\bq + \frac{1}{a}\bq^\perp\, .\label{eq:a_b_solutions_b}
	\end{align}
\end{subequations}
In light of \eqref{eq:n_y_setting}, \eqref{eq:q_0=n_0}, and \eqref{eq:deformation_gradient}, we also see that \eqref{eq:a_b_solutions_a} implies that 
\begin{equation}
	\label{eq:q=n}
	\bq=\n.
\end{equation}

\subsection{Developability of the deformed configuration}
The representation of the deformed configuration assumed in \eqref{eq:y_representation} is that of a ruled surface.
However, since there exist ruled surfaces that are not developable, it needs to be determined whether or not \eqref{eq:y_representation} represents a developable surface.
We show in this section that as a consequence of  \eqref{eq:general_solution}, the surface representation \eqref{eq:y_representation} is indeed developable.
We also show that the normal to the surface coincides with the corresponding $\bd_2(s)$ along a given generator.

Whenever \eqref{eq:inextensibility} is satisfied, the unit normal to the surface is defined as $\bnu = \ba\times\bb$.
Using \eqref{eq:a_b_solutions}, one immediately sees that $\bnu$ can be written as,
\begin{align}
\bnu(s,t)=\bq(s)\times\bq^\perp(s) = \bd_2(s)\, ,\label{eq:normal}
\end{align}
\nigh{and so it turns out to be independent of $t$.}

Next we show that the Gaussian curvature $K=0$, confirming that the mapping under consideration results in a developable configuration. 
Resorting again to a generic curve $(s(\xi),t(\xi))$ we can write,
\begin{equation}\label{eq:pre_curvature}
\dot{\bnu}=\dot{\bd}_2=(u_1\bd_3-u_3\bd_1)\dot{s}=(\nablas\bnu)\dot{\by}\, ,
\end{equation}
where \nigh{again a superimposed dot denotes differentiation with respect to $\xi$ and }$\nablas$ is the \nigh{surface} gradient on the deformed surface.  By \eqref{eq:y_dot}, we see that the curvature tensor (which is symmetric) must be of the form 
\begin{align}
\nablas\bnu=\sigma\bq^\perp\otimes\bq^\perp\, ,\label{eq:curvature_tensor}
\end{align}
\nigh{an expression already obtained in \cite{chen:representation_2015} (see their equation (73)),}
implying that the Gaussian curvature of the deformed configuration vanishes identically. 
Inserting \eqref{eq:curvature_tensor} into \eqref{eq:pre_curvature} and by use of \eqref{eq:consequence}, $\sigma$ can be written as,
\begin{equation}\label{eq:curvature}
 \sigma =\frac{a^2u_1v_3}{\cos\alpha_0\left[\cos\alpha_0+t(\partial_s\alpha_0-\omega_2)\right]},
\end{equation} 
which completes the determination of the curvature tensor.

\subsection{Isometric limit}
We now consider the limit in which the deformation map $\by$ is an isometry,\footnote{Here is where our development crosses path with the vast literature on the equilibrium shapes of an elastic M\"obius band, represented, as suggested in \cite{chen:moebius}, by the following (incomplete) list of studies \cite{schwarz:dark,mahadevan:shape,randrup:sides,hangan:elastic,sabitov:isometric,starostin:shape,starostin:equilibrium_2007,kurono:flat,chubelaschwili:elastic,naokawa:extrinsically,dias:non-linear,kirby:gamma-limit,starostin:equilibrium_2015,shen:geometrically,scholtes:variational}.} and compare the results with some of the standard strip models in the literature such as Wunderlich's \cite{wunderlich:abwickelbares} and Sadowsky's \cite{sadowsky:elementarer} models.\footnote{For the latter, see also \cite{sadowsky:differentialgleichungen,sadowsky:theorie}.}
The Cauchy-Green tensor of the deformation $\by$ as characterized by \eqref{eq:general_solution} can be written as,
\begin{equation}\label{eq:C}
\bC = a^2 \bq_0\otimes\bq_0 + (\ba\cdot\bb)(\bq_0\otimes\bq_0^\perp + \bq_0^\perp\otimes\bq_0) + b^2 \bq_0^\perp\otimes\bq_0^\perp\, ,
\end{equation}
where $a=|\ba|$ and $b=|\bb|$. 
It is then a simple matter to see, by use of  \eqref{eq:general_solution_2} and \eqref{eq:a_b_solutions}, that $\bC = \bI_2$, where $\bI_2$ is the two-dimensional identity, if and only if
\begin{align}
a=1\,,\qquad v_3 =1\, ,\label{eq:isometry_conditions}
\end{align}
\nigh{in accord with equations (18) and (19) of \cite{chen:representation_2015},}
which further implies from \eqref{eq:general_solution} that  $\alpha = \alpha_0$ and $u_2 = \omega_2$.

In this case, we can further characterize the generatrix $\br$ of the deformation $\by$ in \eqref{eq:y_representation}. 
Letting $\kappa>0$ and $\tau$ denote the curvature and torsion of $\br$, for $a=1$ we can also represent $\bq$ as
\begin{equation}\label{eq:q_rewritten}
\bq=\cos\alpha_0\bd_1+\sin\alpha_0\Tangent,
\end{equation} 
where $\Tangent:=\br'$ is the unit tangent to the directrix $\br$. 
By computing $\bq'$ from this equation and combining the result with \eqref{eq:consequence}, we see that if $\omega_2 = 0$ and $\alpha_0\neq0$ then 
\begin{equation}
\label{eq:dichotomy}
\bd_2=\pm\Normal\quad\text{and correspondingly}\quad u_1=\mp\kappa,
\end{equation}
where $\Normal$ is the principal normal of $\br$. 
By use of the Frenet-Serret equations, we easily convert this dichotomy into the following alternative.
\begin{subequations}	\label{eq:alternative}
	\begin{align}
	\text{either}&\quad 	\bd_1=-\Binormal,\quad\bd_2=\Normal,\qquad u_1=-\kappa,\quad u_3=-\tau,\\
	\text{or}&\quad \bd_1=\Binormal,\quad\bd_2=-\Normal,\qquad u_1=\kappa,\quad u_3=-\tau,
	\end{align}
\end{subequations}
\nigh{where $\Binormal:=\Tangent\times\Normal$ is the binormal unit vector.}
Correspondingly,
\begin{equation}
\label{eq:torsion}
\tau=\pm\kappa\tan\alpha_0.
\end{equation}
Combining \eqref{eq:q_rewritten}, \eqref{eq:alternative}, and  \eqref{eq:torsion}, we can write
\begin{equation}
\label{eq:q_isometry}
\bq=\cos\alpha_0(\pm\Binormal+\tan\alpha_0\Tangent)=\cos\alpha_0\left(\pm\Binormal\mp\frac{\tau}{\kappa}\Tangent\right),
\end{equation}
which, once inserted into \eqref{eq:y_representation}, delivers a formula that differs only by a sign\footnote{\nigh{At variance with \cite{chen:representation} (see their equation (5.3)), we adopted a convention for the sign of $\tau$ according to which the third Frenet-Serret equation reads as $\Binormal'=\tau\Normal$}.} from equation (1.6) of \cite{chen:representation}.

\nigh{In our approach, a material frame is preferred to the Frenet-Serret frame to describe the orientation in space of the ribbon, with the advantage of being also applicable at points where $\kappa=0$.}
\section{Ribbon energy}\label{sec:energy}
In this section we carry out the further dimension reduction of  the (scaled) per area energy $\tilde{f}_e$ of a thin sheet of NPN in \eqref{eq:free_energy_functional} to obtain a one-dimensional energy for a ribbon. In \eqref{eq:free_energy_functional}, we set $f_s=hf_1$ and $f_b=h^3f_3$.
Furthermore, it results from the kinematic analysis performed in the preceding section that $f_1$ and $f_3$ are given by 
\begin{subequations}
	\begin{align}
	f_1 &= \frac{2}{1+S}\left[\tr\bC + S_0 \bq_0\cdot\bC\bq_0 + \frac{S}{\bq_0\cdot\bC\bq_0}\right]\, ,\label{eq:f1}\\
	f_3&=\frac{8}{3}\sigma^2\, .\label{eq:f3}
	\end{align}
\end{subequations}
The expression for $f_3$ in \eqref{eq:f3} embodies a significant simplification of equation \eqref{eq:f_b} due to the fact that $K=0$ and $\kappa_n=0$ in our current setting. Also, $\sigma = 2H$ where $H$ is the mean curvature.
The expression on the right in \eqref{eq:f1} corresponds to equation \eqref{eq:f_s}.
Also, it is worth recalling that in writing \eqref{eq:f1}, the unit vector $\bq_0$ has been identified with the director field $\n_0$ imprinted on the reference configuration, as postulated  in \eqref{eq:q_0=n_0}. 
While $S_0$ is determined at the time of cross-linking,  we recall that $S$ can be affected by external stimuli (such as light and heat); it is precisely the difference that can be induced (by external agents) between $S_0$ and $S$ that drives the spontaneous deformation of a NPN sheet. 
 
We first compute the area element $dA$ of the reference configuration $\ribbon$ which we would require for later use for dimensional reduction. The area element is given by,
\begin{align}
	dA=\lvert\partial_s\bx\times\partial_t\bx\rvert ds \,dt=\left[\cos\alpha_0 + t(\partial_s\alpha_0-\omega_2)\right]ds\, dt\, .\label{eq:area_element}
\end{align}
The (scaled) total elastic free energy $\free$ of the full surface $\ribbon$ in \eqref{eq:free_energy_functional} can then be written as, 
\begin{equation}\label{eq:elastic_free_energy}
\free[\y] =\int_0^L \int_{t_-}^{t_+}\! (hf_1+h^3f_3)[\cos\alpha_0 + t\left(\partial_s\alpha - \omega_2\right)]dt\,ds =\int_0^L\!  f\, ds,
\end{equation}
where $L$ is the length of the directrix $\br_0$ in the reference configuration and $f$ represents the (scaled) energy  per unit arc-length of the directrix $\br_0$.
After some lengthy, but not difficult computations, we obtain the following representation of the reduced energy,
\begin{align}
f= F_0\ln\left(\frac{\cos\alpha_0 + t_+ M}{\cos\alpha_0 + t_- M}\right) +F_1 (t_+ - t_-)+F_2 (t_+^2 - t_-^2)\,,\label{eq:F}
\end{align} 
where 
\begin{subequations}
	\begin{align}
	F_0 & = \left[\frac{8 a^4 h^3 u_1^2 v_3^2}{3 M \cos\alpha_0^2}+\frac{2h\left(VM - \partial_s a\cos\alpha_0\right)^2}{(1+s)M^3}\right]\, ,\label{eq:F_0}\\
	F_1 & = \left[\frac{\cos\alpha_0}{a^2}+\frac{1+S_0}{1+S}a^2\cos\alpha_0+\frac{\partial_s a (2VM - \partial_sa\cos\alpha_0)}{(1+S) M^2}\right] 2h \, ,\label{eq:F_1}\\
	F_2 &= \left[\frac{M}{a^2} + \frac{1+S_0}{1+S}a^2 M+\frac{(\partial_s a)^2}{(1+S)M}\right]h\, ,\label{eq:F_2}\\
	V & = v_3\sin\alpha - a \sin\alpha_0 \, ,\label{eq:A}\\
	M &= \partial_s\alpha_0 - \omega_2\,.\label{eq:M}
	\end{align}
\end{subequations}
Details of this derivation are presented in Appendix \ref{app:dimension_reduction}. Although $a$ has a direct geometric meaning, which can be read off from \eqref{eq:C}, we find it convenient to express it via \eqref{eq:general_solution_1} in terms of two other measures of deformation, namely, $v_3$ and the angle $\alpha$ that $\bq$ makes with $\dv_1$,
\begin{equation}
\label{eq:a_alpha}
a=\frac{\cos\alpha_0}{v_3\cos\alpha}.
\end{equation}
In this way, equation \eqref{eq:general_solution_2} is effectively incorporated.

Since $\alpha_0$ is known, by use of \eqref{eq:a_alpha}, for prescribed $S_0$ and $S$, $\free$ becomes a functional in the triple of functions $(v_3,\alpha,u_1)$ subject to the boundary conditions appropriate to the specific problem at hand. For the minimizing triple, both \eqref{eq:general_solution_3} and \eqref{eq:general_solution_4} then also provide $u_2$ and $u_3$, thus completing the description of the shape acquired by the actuated NPN ribbon. 

Next we reduce the general energy function\eqref{eq:F} to a simple case of rectangular geometry, and present an example where the ribbon undergoes in-plane serpentine deformations.

\section{Rectangular geometry}\label{sec:geometry}
\begin{figure}[h]
	\centering
	\includegraphics[width=0.8\textwidth]{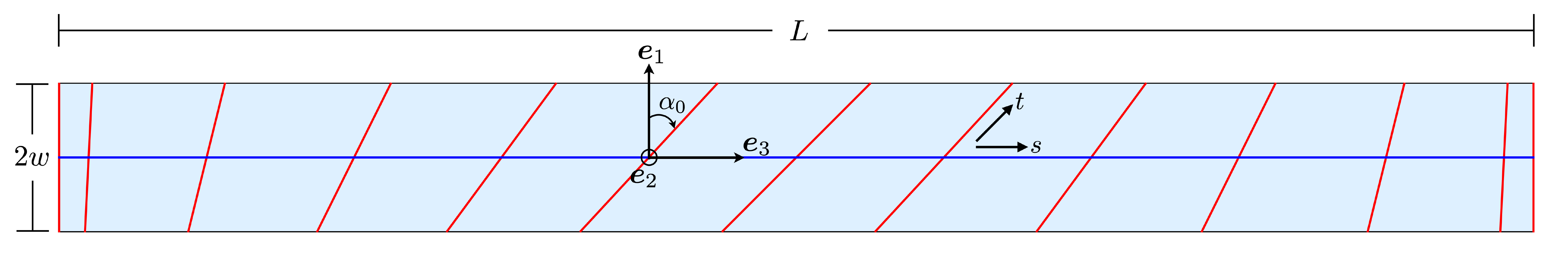}
	\caption{Natural configuration of a rectangular ribbon.}
	\label{fig:rectangle_schematics}
\end{figure}

In this section, we consider a special NPN ribbon whose stress free natural configuration is a rectangle of width $2w$ and length $L$ (see Fig. \ref{fig:rectangle_schematics}).  We shall take the centerline $\br_0$ to lie along the $\e_3$ axis, so that $s=x_3$ and $\omega_2=0$. We further let $\alpha_0$ be a smooth function such that 
\begin{equation}
\label{eq:alpha_0_boundary_conditions}
\alpha_0(0)=\alpha_0(L)=0,
\end{equation}
so that the short sides of the ribbon are both along $\e_1$. With such a geometric choice,
\begin{equation}\label{eq:t's}
t_+=-t_-=\frac{w}{\cos\alpha_0}\,.
\end{equation}	
Thus, the contributions to $f$ in \eqref{eq:F} that are quadratic in $w$ vanish identically, and we obtain,
\begin{align}
f = F_0\ln\left(\frac{1 + w \,\partial_s\alpha_0 \sec^2\alpha_0}{1 - w\, \partial_s\alpha_0 \sec^2\alpha_0}\right) + 2F_1 w \sec\alpha_0\, ,\label{eq:rec_energy_1}
\end{align}
where we have divided the numerator and the denominator inside the $\ln$ by $\cos\alpha_0$.

We briefly digress here to compare the above expression for a rectangular ribbon of finite width with that of Wunderlich's model, which considers isometric deformations. We set $a=1$, $v_3=1$ and $\alpha = \alpha_0$ as per the conclusions reached in \eqref{eq:isometry_conditions}, along with $S=S_0=0$.
Under these conditions, $M=\partial_s\alpha$ and $V = 0$ from \eqref{eq:A} and \eqref{eq:M}.
Substituting these in \eqref{eq:rec_energy_1}, and identifying $\tan\alpha$ with $\eta$ and $u_1^2 = \kappa^2$, the energy density above reduces to,
\begin{align}
f = \frac{8h^3\kappa^2}{3\eta'}\left(1+\eta^2\right)^2\ln\left(\frac{1+w\eta'}{1-w\eta'}\right) + 4hw\, ,\label{eq:wunderlich}
\end{align}
which differs from the Wunderlich's energy for a rectangular strip of finite width by an inessential additive constant.
It is then easy to see that \nigh{at the lowest order in $w$}, the above expression reduces to Sadowsky's energy for rectangular ribbons with small width.\footnote{Leaving aside the somewhat exotic generalization proposed in \cite{freddi:corrected}.}

Returning to our main discussion,  \nigh{at the lowest order in $w$} in expression \eqref{eq:rec_energy_1}, and using \eqref{eq:a_alpha}, \eqref{eq:A} and \eqref{eq:M} we obtain the following energy functional,
\begin{equation}
\free[\y]=4w\int_0^L\left[\frac43h^3\frac{u_1^2}{v_3^2\cos^4\alpha}+h\left(\frac{v_3^2\cos^2\alpha}{\cos^2\alpha_0}+\frac{1+S_0}{1+S}\frac{\cos^2\alpha_0}{v_3^2\cos^2\alpha}+\frac{(v_3^2\cos\alpha\sin\alpha-\cos\alpha_0\sin\alpha_0)^2}{(1+S)v_3^2\cos^2\alpha_0\cos^2\alpha}\right)\right] ds,\label{eq:elastic_free_energy_rewritten}
\end{equation}
which, in particular, is independent of $\partial_sa$. 

The boundary conditions that we now consider will identify one special case, where the deformed ribbon is expected to be in the reference plane, while taking on a serpentine configuration.

\subsection{Serpentining ribbon}\label{sec:serpentining}
Now, we fix the end-point at $s=0$ of the ribbon and clamp the corresponding edge, so that frames $\framed$ and $\framee$ coincide for $s=0$, while both the end-point at $s=L$ and the corresponding edge are left free. 
Clearly, $u_1=0$ is compatible with all these boundary conditions and, furthermore, minimizes $\free$ in \eqref{eq:elastic_free_energy_rewritten} for any $\alpha_0$, as $u_1$ is independent of both $v_3$ and $\alpha$. 
Consequently, by \eqref{eq:general_solution_3} and \eqref{eq:general_solution_4}, we conclude that also $u_3\equiv0$ and
\begin{equation}
\label{eq:theta_prime}
\partial_s\vartheta=\partial_s\alpha-\frac{v_3^2\cos^2\alpha}{\cos^2\alpha_0}\partial_s\alpha_0,
\end{equation}
where we have used the relation $u_2=\partial_s\vartheta$, with $\vartheta$ being the angle between $\bd_1$ and $\be_1$. 
Expression \eqref{eq:theta_prime} is to be integrated, once both $\alpha$  and $v_3$ are determined, with the initial condition $\vartheta(0)=0$.

Since neither $\partial_s\alpha$ nor  $\partial_s v_3$ feature in $\free$, the optimal $\alpha$ and $v_3$ are the minimizers of the integrand of $\free$. 
Despite the apparent complexity of the task, finding these minimizers is boring, but not difficult. 
Only  one such minimizing pair actually exists and is delivered explicitly by
\begin{subequations}\label{eq:minimizing_pair}
	\begin{align}
	\alpha&=\arctan(\mu\tan\alpha_0)\,,\label{eq:minimizing_pair_1}\\
	v_3&=\frac{\sqrt{(S_0-S)\cos^2\alpha_0+S+1}}{\sqrt[4]{(S_0+1)(S+1)}}\,,\label{eq:minimizing_pair_2}	
	\end{align}
\end{subequations}
where, for brevity, we have set
\begin{equation}\label{eq:mu_definition}
\mu:=\sqrt{\frac{S+1}{S_0+1}}.	
\end{equation}
It follows from \eqref{eq:minimizing_pair} that $a\equiv\sqrt{\mu}$ and
\begin{equation}
\label{eq:theta_solution}
\vartheta=\arctan(\mu\tan\alpha_0)-\frac{\alpha_0}{\mu},
\end{equation}
which determines the deformed shape of the ribbon's centerline through the following quadratures,
\begin{equation}
\label{eq:shape_quadratures}
x_3^\ast(s)=\int_0^s\!v_3(\xi)\cos\vartheta(\xi)d\xi,\quad x_1^\ast(s)=\int_0^s\!v_3(\xi)\sin\vartheta(\xi)d\xi\,,
\end{equation}
while its total length $L^\ast$ is given by
\begin{equation}
\label{eq:L_star}
L^\ast=\int_0^L\!v_3(s)ds.
\end{equation}
\begin{figure}[h!]
	\centering
	\begin{subfigure}[c]{.48\linewidth}
		\centering
		\includegraphics[width=\linewidth]{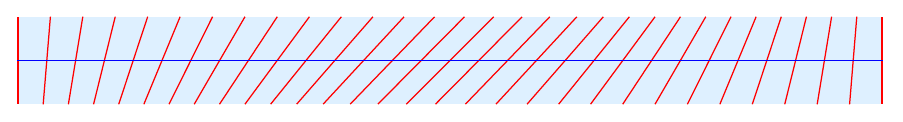}
		\caption{$n=1$}
	\end{subfigure} 
	\begin{subfigure}[c]{.48\linewidth}
		\centering
		\includegraphics[width=.94\linewidth]{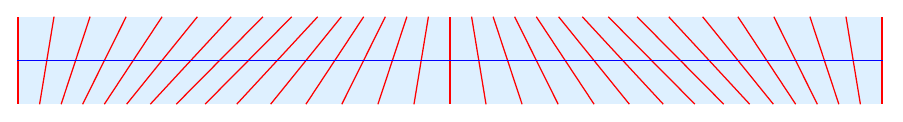}
		\caption{$n=2$}
	\end{subfigure}
	\begin{subfigure}[c]{.48\linewidth}
		\centering
		\includegraphics[width=.92\linewidth]{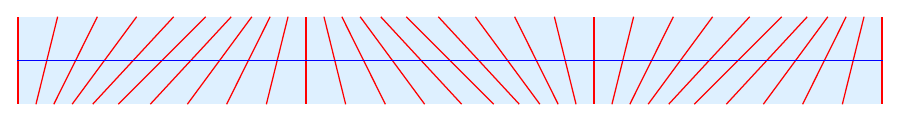}
		\caption{$n=3$}
	\end{subfigure}
	\caption{Representation of $\alpha_0$ in \eqref{eq:alpha_0_illustration} for a straight undeformed ribbon. The bends in the imprinted director $\bq_0$ will result in bends of the ribbon's centerline upon activation. The width of the ribbons is exaggerated for clarity of presentation.}
	\label{fig:ribbons}
\end{figure}
As an illustration, we determine the deformed centerline of the ribbon when the imprinted nematic director $\bq_0$ is described by 
\begin{equation}\label{eq:alpha_0_illustration}
\alpha_0(s)=\frac\pi4\sin\left(\frac{n\pi s}{L}\right),
\end{equation}
\begin{figure}[h]
	\centering
	\begin{subfigure}[c]{.49\linewidth}
		\centering
		\includegraphics[width=\linewidth]{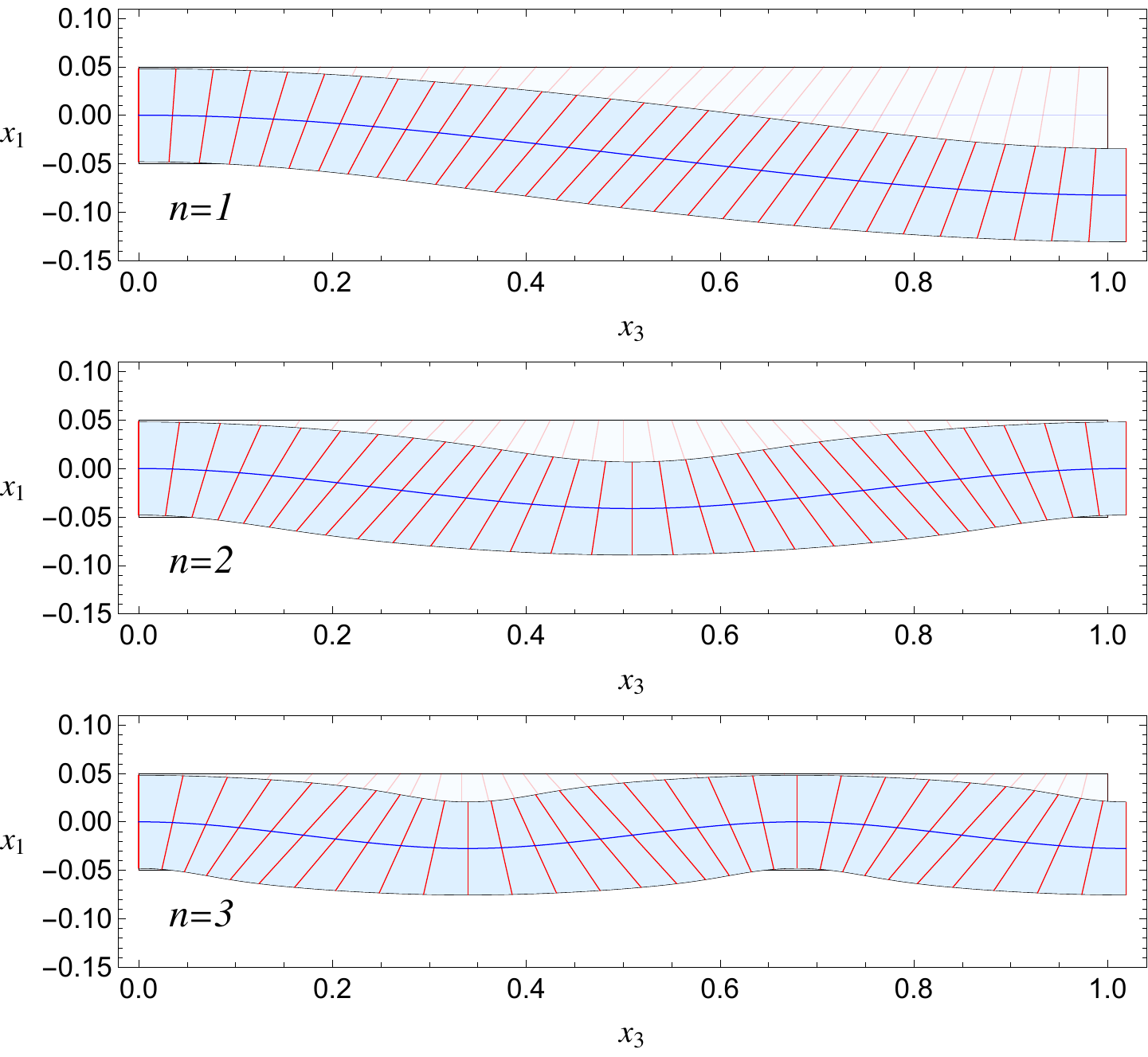}
		\caption{$S_0=0.5$, $S=0.25$, $L^\ast/L\doteq1.023$.}
	\end{subfigure} 
	\begin{subfigure}[c]{.49\linewidth}
		\centering
		\includegraphics[width=\linewidth]{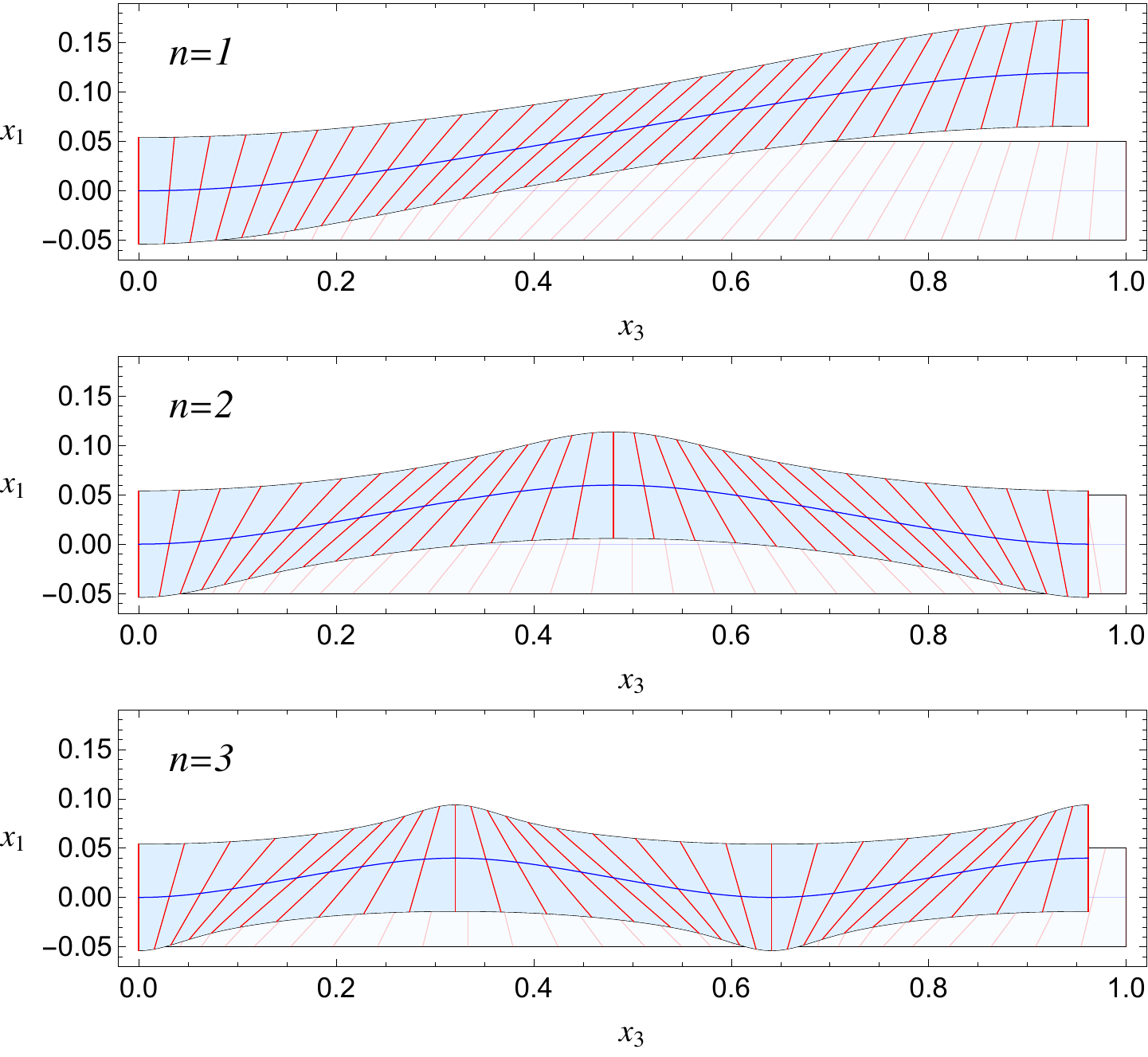}
		\caption{$S_0=0.5$, $S=1.0$, $L^\ast/L\doteq0.97$.}
	\end{subfigure}
	\caption{Serpentining shapes of the ribbons depicted in Fig.~\ref{fig:ribbons} upon activation. Lengths are scaled to the length $L$ of the undeformed ribbon. $L^\ast$, the length of the deformed centerline, is the same for all $n=1,2,3$. The number of bends equals $n$.}
	\label{fig:centerlines}
\end{figure}
where $n$ is an integer. Fig.\ref{fig:ribbons} shows the decoration of the undeformed ribbon for $n=1,2,3$, while the corresponding deformed centerline is depicted in Fig.~\ref{fig:centerlines} for both $S>S_0$ and $S<S_0$; they represent serpentining ribbons with $n$ bends swerving on opposite sides of the undeformed ribbon.
The planar deformed configurations corresponding to the three patterns of the imprinted director are shown in Fig. \ref{fig:centerlines}.
The planar bending of the centerline is a vestige of the anisotropic stretching of the NPN sheet the ribbon was carved from.

\section{Conclusion}\label{sec:conclusion}
Motivated by a growing body of studies, both experimental and theoretical in nature, about the photoactivation of nematic elastomers (see, for example, \cite{goriely:rod}), we derived the elastic free-energy functional whose minimizers are the spontaneous equilibrium shapes of a NPN ribbon activated by a mismatch in molecular order induced by an external agent. Our starting point for this derivation was the plate theory that had been developed in \cite{ozenda:blend} from the \emph{neo-classical} formula for the energy density of nematic elastomers in three-dimensional space.  

The explicit expression for the elastic energy-density (per unit length of the undeformed ribbon's centerline) is admittedly rather involved, but only because it is very general: the ribbon's centerline can be arbitrarily stretched, although the ribbon is itself an inextensible surface, and virtually no geometric limitations are imposed on the deformation of the ribbon. 

The elastic free-energy functional we obtained was compared to the classical Sadowsky's and Wunderlich's functionals, valid only for isometric deformations, which represent limiting cases of deformations allowed in our model. This is where our development intersected the vast literature on the equilibria of a M\"obius band, but with a major, essential difference: in our theory the straight lines that decompose the developable surfaces representing the ribbon in both reference and current configurations are material in nature, as they are determined by the \emph{imprinted} nematic director.

To put our theory to the test, we considered the case of a rectangular ribbon, on which modulated director fields were imprinted at the time of cross-linking. The boundary conditions were compatible with spontaneous in-plane deformations, which we determined explicitly, showing how the serpentining shape of the deformed ribbon resonates with modulated director fields. Clearly, this is a quite simple application of our theory, where its potential of describing out-of-plane activated shapes of a ribbon is not fully deployed. To accomplish this latter task, which is presently being undertaken, we \nigh{shall adapt to} the present setting the theory put forward in \cite{hornung:euler}, \nigh{as we found intractable the Euler-Lagrange equations to the functional \eqref{eq:elastic_free_energy_rewritten} derived from the classical formulae for the strains in terms of the representation of the director frame $\framed$ in a fixed, reference frame (see p.\,301 of \cite{antman:nonlinear}).}

In this paper, and in many others, the effect of an external stimulus is described in a simple, phenomenological way: we assume that it can be reduced to a uniform change in a scalar order parameter, which drives the ribbon out of equilibrium. This approach might not be sufficient to describe the details of photoactivation, which was the original inspiration of this work. Light is absorbed in solids in a way that depends on how it propagates, as described, for example, by Beer's law \cite{beer:bestimmung} (see also Chapt. 1 of \cite{fox:otptical}), which here becomes more stringent, as in general the activable mesogens are just a fraction  of polymer strands. Implementing these details in our ribbon theory might make it applicable to real-life experiments, where the induced elastic behaviour also depends, possibly quite dramatically,  on the direction of propagation of light (see, for example, \cite{camacho-lopez:fast,liu:light}).

\section*{Acknowledgments}
H.S. acknowledges partial support by Swiss National Science Foundation Grant $200020\_182184$ to J.H. Maddocks.

\appendix

\section{Step-length and nematic order tensors}\label{app:step_tensor}
In this appendix, we relate explicitly the geometric activation parameter $S$ that features in the step-length tensor $\step$ in \eqref{eq:step_tensor} to the Maier-Saupe scalar order parameter $Q$ for the rod-like molecules that constitute the polymer strands in the network. As $Q$ can be altered in response to a temperature change, so can also be $S$.

We assume that all polymer strands are constituted by $N$ rigid rods, each of length $a$,\footnote{The notation ``$a$'' here is unrelated to, and should not be confused with, the scalar function $a(s)$ introduced in equation \eqref{eq:general_solution_1} in the main text.} freely jointed, one to the next, at their end-points, so as to form a flexible chain. The \emph{span vector} $\bR$ of a chain is given by
\begin{equation}
	\label{eq:span_vector}
	\bR=a\sum_{i=1}^N\bu_i,
\end{equation}
where $\bu_i$ is the unit vector oriented along the $i$-th rod. The step-length tensor $\step$ is defined as
\begin{equation}
	\label{eq:step_tensor_definition}
	\step:=\frac{3}{Na}\ave{\bR\otimes\bR},
\end{equation}
where the brackets $\ave{\cdots}$ denote local ensemble average. Since all rods are equally free to rotate in all directions, $\ave{\bu_i}=\zero$, for all $i$, which implies $\ave{\bR}=\zero$, so that $\step$ represents the lowest measure of geometric order in the system. The mutual independence of the rods then readily leads us to conclude that 
\begin{equation}
	\label{eq:step_tensor_representation}
	\step=\frac{3Na^2}{Na}\ave{\bu\otimes\bu}=3a\left(\bQ+\frac13\bI\right),
\end{equation}
where $\bI$ is the identity tensor and
\begin{equation}
	\label{eq:Q_definition}
	\bQ:=\ave{\bu\otimes\bu-\frac13\bI}
\end{equation}
is the nematic \emph{quadrupolar} order tensor. Assuming that $\bQ$ is uniaxial about the nematic director $\bn$, we can write
\begin{equation}
	\label{eq:Q_representation}
	\bQ=Q\left(\bn\otimes\bn-\frac13\bI\right).
\end{equation}
Here $Q$ is the scalar order parameter, which can also be expressed in the form
\begin{equation}\label{eq:Q_order_parameter}
	Q=\frac32\left(\ave{(\bu\cdot\bn)^2}-\frac13\right)=\ave{P_2(\cos^2\vartheta)},
\end{equation}	
where $P_2$ is the Legendre polynomial of degree $2$ and $\vartheta$ is the angle that the unit vector $\bu$ makes with the director $\bn$.

Clearly, $Q$ ranges in the interval $[-\frac12,1]$, the upper bound of which is attained (in the unphysical case) when all rods are aligned along $\bn$, while the lower bound corresponds to the case where all rods are randomly distributed in the plane orthogonal to $\bn$. When $Q$ vanishes so also does $\bQ$ and $\bL$ reduces to $a\bI$, which represents the isotropic distribution of rods.\footnote{The strand is then statistically equivalent to a globule of radius $a$, thus fully justifying the scaling adopted in the definition \eqref{eq:step_tensor_definition}.}

By use of \eqref{eq:Q_representation} in \eqref{eq:step_tensor_representation}, we arrive at (see also \cite{corbett:nonlinear,corbett:polarization})
\begin{equation}
	\label{eq:step_tensor_repeated}
	\bL=A(\bI+S\bn\otimes\bn),
\end{equation}
which is the same tensor $\bL$ in \eqref{eq:step_tensor}, where we can now make the following identifications,
\begin{equation}
	\label{eq:identifications}
	A=a(1-Q)\quad\text{and}\quad S=\frac{3Q}{1-Q}.
\end{equation}
Since $S$ is an increasing function of $Q$, an increase in order, such as that produced by a decrease in temperature, delivers an increase in $S$. Contrariwise, an increase in order determines a decrease in $A$.

Alternatively, $\step$ has also been represented as follows \cite{verwey:elastic},
\begin{equation}
	\label{eq:step_tensor_alternative}
	\step=\ell_\perp\Big(\bI+(r-1)\bn\otimes\bn\Big),
\end{equation}
where $\ell_\perp$ is the principal chain step length in the direction orthogonal to $\bn$ and $\ell_\parallel:=r\ell_\perp$ is the principal chain step length in the direction parallel to $\bn$. Comparing \eqref{eq:step_tensor_alternative} and \eqref{eq:step_tensor_repeated}, we easily identify the relations
\begin{equation}
	\label{eq:relations}
	A=\ell_\perp,\quad S=r-1.
\end{equation}

In the main text of the paper, two step-length tensors have played a role: the one associated with the nematic order established at the time of cross-linking, and the one associated with the nematic order induced by the activation process at work; we have denoted them as $\step_0$ and $\step$, with nematic directors $\bn_0=\bq_0$ and $\bn=\bq$ and scalar order parameters $S_0$ and $S$, respectively. These are related via \eqref{eq:relations}  to the standard scalar order parameters $Q_0$ and $Q$ and to the corresponding amplitudes $A_0$ and $A$. Following \cite{corbett:nonlinear,corbett:polarization,ozenda:blend}, one then realizes that the \nigh{physical constant on the right side of \eqref{eq:F_definition} can also be written as}
\begin{equation}
	\label{eq_k_representation}
	\nigh{\frac12k\frac{A_0}{A}}=\frac12n_sk_BT\frac{A_0}{A}=\frac12n_sk_BT\frac{1-Q_0}{1-Q},
\end{equation} 
where $n_s$ is the number of polymer strands per unit volume, $k_B$ is the Boltzmann constant and $T$ the absolute temperature.

\section{Derivation of the general solution}\label{app:general_solution}
Here we present the details of the computations that result in the general solution \eqref{eq:general_solution} from the inextensibility constraint \eqref{eq:inextensibility}.
Multiplying both sides of equation \eqref{eq:inextensibility} with the denominator on the left, and using \eqref{eq:q} we obtain,
\begin{gather}
\left(\partial_t\psi\right)^2\left\{v_3^2\cos^2\alpha - 2 v_3\cos\alpha(u_2-\partial_s\alpha)\psi + \left[(u_2 - \partial_s\alpha)^2 - (u_1\sin\alpha-u_3\cos\alpha)^2\right]\psi^2\right\} \nonumber\\ =\cos^2\alpha_0 
                   																									  + 2\cos\alpha_0(u_2-\partial_s\alpha)t
                   																									 +  (u_2-\partial_s\alpha)^2 t^2 \, .\label{eq:inextensibility_expanded}
\end{gather}
Since the right hand side is a second degree polynomial in $t$, and the only $t$ dependent function on the left is $\psi$, one can conclude from \eqref{eq:inextensibility_expanded} that $\psi$ must be linear in $t$. 
Combining this with the requirement that $\psi(s,0) = 0$, leads to equation \eqref{eq:general_solution_1}.
Thereafter, equating the zeroth, first, and second order terms in $t$ on both sides, one arrives at equations \eqref{eq:general_solution_2}, \eqref{eq:general_solution_3}, and \eqref{eq:general_solution_4} respectively.

\section{Details of energy reduction}\label{app:dimension_reduction}
Here we provide details of some computations needed to obtain the  reduced energy density $f$ stated in \eqref{eq:F}.
From the last equality in \eqref{eq:elastic_free_energy}, the reduced energy $f$ (per unit $k$) can be written as,
\begin{align}
f = \int_{t_-}^{t_+}\!\! (h f_1 + h^3f_3)\left[\cos\alpha_0 + t(\partial_s\alpha - \omega_2)\right]dt\, ,\label{eq:F_simplified}
\end{align}
where $f_1$ and $f_3$ are given by equations \eqref{eq:f1} and \eqref{eq:f3}.
Using \eqref{eq:q0} and \eqref{eq:C}, the expression for $f_1$ can be further simplified to
\begin{align}
f_1 &= \frac{2}{1+S}\left[a^2(1+S_0) + \frac{S}{a^2} + b^2\right]\, ,\label{eq:f1_simplified}
\end{align}
where $a$ and $b$ are the magnitudes of the vectors $\ba$ and $\bb$ from \eqref{eq:a} and \eqref{eq:b}. 
While $a$ is a basic unknown, $b$ can be read off from \eqref{eq:b} as
\begin{align}
b = \sqrt{\left(\frac{v_3\sin\alpha - a\sin\alpha_0 + t\partial_s a}{\cos\alpha_0 + t(\partial_s\alpha_0 - \omega_2)}\right)^2 + \frac{1}{a^2}}\, .\label{eq:b_mag}
\end{align}
Upon substituting for $f_1$ and $f_3$ from \eqref{eq:f1_simplified} and \eqref{eq:f3}, and using \eqref{eq:curvature} and \eqref{eq:b_mag} thereafter, integral \eqref{eq:F_simplified} can be rearranged in the following form,
\begin{align}
\begin{split}
f =&\int_{t_-}^{t_+}\!\!\left[\frac{8a^4h^3u_1^2v_3^2}{3M\cos\alpha_0^2}+\frac{2h(VM - \partial_s a \cos\alpha_0)^2}{(1+S)M^3}\right]\frac{M}{\cos\alpha_0 + t M}\, dt\\
 & +\int_{t_-}^{t_+}\!\! \left[\frac{\cos\alpha_0}{a^2} + \frac{1+S_0}{1+S}a^2\cos\alpha_0 + \frac{\partial_s a (2VM - \partial_s a\cos\alpha_0)}{(1+S)M^2}\right]2h\, dt\\
 &+\int_{t_-}^{t_+}\!\!\left[\frac{M}{a^2} + \frac{1+S_0}{1+S}a^2 M + \frac{(\partial_s a)^2}{(1+S)M}\right]2ht \,dt \, ,
\end{split}\label{eq:F_integrand}
\end{align}
where $V$ and $M$ are defined in \eqref{eq:A} and \eqref{eq:M}.
A straightforward integration of the three integrals above delivers expression \eqref{eq:F}.


\begin{thebibliography}{83}%
	\makeatletter
	\providecommand \@ifxundefined [1]{%
		\@ifx{#1\undefined}
	}%
	\providecommand \@ifnum [1]{%
		\ifnum #1\expandafter \@firstoftwo
		\else \expandafter \@secondoftwo
		\fi
	}%
	\providecommand \@ifx [1]{%
		\ifx #1\expandafter \@firstoftwo
		\else \expandafter \@secondoftwo
		\fi
	}%
	\providecommand \natexlab [1]{#1}%
	\providecommand \enquote  [1]{``#1''}%
	\providecommand \bibnamefont  [1]{#1}%
	\providecommand \bibfnamefont [1]{#1}%
	\providecommand \citenamefont [1]{#1}%
	\providecommand \href@noop [0]{\@secondoftwo}%
	\providecommand \href [0]{\begingroup \@sanitize@url \@href}%
	\providecommand \@href[1]{\@@startlink{#1}\@@href}%
	\providecommand \@@href[1]{\endgroup#1\@@endlink}%
	\providecommand \@sanitize@url [0]{\catcode `\\12\catcode `\$12\catcode
		`\&12\catcode `\#12\catcode `\^12\catcode `\_12\catcode `\%12\relax}%
	\providecommand \@@startlink[1]{}%
	\providecommand \@@endlink[0]{}%
	\providecommand \url  [0]{\begingroup\@sanitize@url \@url }%
	\providecommand \@url [1]{\endgroup\@href {#1}{\urlprefix }}%
	\providecommand \urlprefix  [0]{URL }%
	\providecommand \Eprint [0]{\href }%
	\providecommand \doibase [0]{https://doi.org/}%
	\providecommand \selectlanguage [0]{\@gobble}%
	\providecommand \bibinfo  [0]{\@secondoftwo}%
	\providecommand \bibfield  [0]{\@secondoftwo}%
	\providecommand \translation [1]{[#1]}%
	\providecommand \BibitemOpen [0]{}%
	\providecommand \bibitemStop [0]{}%
	\providecommand \bibitemNoStop [0]{.\EOS\space}%
	\providecommand \EOS [0]{\spacefactor3000\relax}%
	\providecommand \BibitemShut  [1]{\csname bibitem#1\endcsname}%
	\let\auto@bib@innerbib\@empty
	\bibitem [{\citenamefont {Warner}\ and\ \citenamefont
		{Terentjev}(2003)}]{warner:liquid}%
	\BibitemOpen
	\bibfield  {author} {\bibinfo {author} {\bibfnamefont {M.}~\bibnamefont
			{Warner}}\ and\ \bibinfo {author} {\bibfnamefont {E.~M.}\ \bibnamefont
			{Terentjev}},\ }\href@noop {} {\emph {\bibinfo {title} {Liquid Crystal
				Elastomers}}},\ \bibinfo {series} {International Series of Monographs on
		Physics}, Vol.\ \bibinfo {volume} {120}\ (\bibinfo  {publisher} {Oxford
		University Press},\ \bibinfo {address} {New York},\ \bibinfo {year}
	{2003})\BibitemShut {NoStop}%
	\bibitem [{\citenamefont {Bladon}\ \emph {et~al.}(1994)\citenamefont {Bladon},
		\citenamefont {Terentjev},\ and\ \citenamefont
		{Warner}}]{blandon:deformation}%
	\BibitemOpen
	\bibfield  {author} {\bibinfo {author} {\bibfnamefont {P.}~\bibnamefont
			{Bladon}}, \bibinfo {author} {\bibfnamefont {E.~M.}\ \bibnamefont
			{Terentjev}},\ and\ \bibinfo {author} {\bibfnamefont {M.}~\bibnamefont
			{Warner}},\ }\bibfield  {title} {\bibinfo {title} {Deformation-induced
			orientational transitions in liquid crystals elastomer},\ }\href
	{https://doi.org/10.1051/jp2:1994100} {\bibfield  {journal} {\bibinfo
			{journal} {J. Phys. II France}\ }\textbf {\bibinfo {volume} {4}},\ \bibinfo
		{pages} {75} (\bibinfo {year} {1994})}\BibitemShut {NoStop}%
	\bibitem [{\citenamefont {Warner}\ \emph {et~al.}(1994)\citenamefont {Warner},
		\citenamefont {Bladon},\ and\ \citenamefont {Terentjev}}]{warner:soft}%
	\BibitemOpen
	\bibfield  {author} {\bibinfo {author} {\bibfnamefont {M.}~\bibnamefont
			{Warner}}, \bibinfo {author} {\bibfnamefont {P.}~\bibnamefont {Bladon}},\
		and\ \bibinfo {author} {\bibfnamefont {E.~M.}\ \bibnamefont {Terentjev}},\
	}\bibfield  {title} {\bibinfo {title} {``{S}oft elasticity''---deformation
			without resistance in liquid crystal elastomers},\ }\href
	{https://doi.org/10.1051/jp2:1994116} {\bibfield  {journal} {\bibinfo
			{journal} {J. Phys. II France}\ }\textbf {\bibinfo {volume} {4}},\ \bibinfo
		{pages} {93} (\bibinfo {year} {1994})}\BibitemShut {NoStop}%
	\bibitem [{\citenamefont {Terentjev}\ \emph {et~al.}(1994)\citenamefont
		{Terentjev}, \citenamefont {Warner},\ and\ \citenamefont
		{Bladon}}]{terentjev:orientation}%
	\BibitemOpen
	\bibfield  {author} {\bibinfo {author} {\bibfnamefont {E.~M.}\ \bibnamefont
			{Terentjev}}, \bibinfo {author} {\bibfnamefont {M.}~\bibnamefont {Warner}},\
		and\ \bibinfo {author} {\bibfnamefont {P.}~\bibnamefont {Bladon}},\
	}\bibfield  {title} {\bibinfo {title} {Orientation of nematic elastomers and
			gels by electric fields},\ }\href {https://doi.org/10.1051/jp2:1994154}
	{\bibfield  {journal} {\bibinfo  {journal} {J. Phys. II France}\ }\textbf
		{\bibinfo {volume} {4}},\ \bibinfo {pages} {667} (\bibinfo {year}
		{1994})}\BibitemShut {NoStop}%
	\bibitem [{\citenamefont {Verwey}\ and\ \citenamefont
		{Warner}(1995{\natexlab{a}})}]{verwey:soft}%
	\BibitemOpen
	\bibfield  {author} {\bibinfo {author} {\bibfnamefont {G.~C.}\ \bibnamefont
			{Verwey}}\ and\ \bibinfo {author} {\bibfnamefont {M.}~\bibnamefont
			{Warner}},\ }\bibfield  {title} {\bibinfo {title} {Soft rubber elasticity},\
	}\href {https://doi.org/10.1021/ma00116a036} {\bibfield  {journal} {\bibinfo
			{journal} {Macromolecules}\ }\textbf {\bibinfo {volume} {28}},\ \bibinfo
		{pages} {4303} (\bibinfo {year} {1995}{\natexlab{a}})}\BibitemShut {NoStop}%
	\bibitem [{\citenamefont {Verwey}\ and\ \citenamefont
		{Warner}(1995{\natexlab{b}})}]{verwey:multistage}%
	\BibitemOpen
	\bibfield  {author} {\bibinfo {author} {\bibfnamefont {G.~C.}\ \bibnamefont
			{Verwey}}\ and\ \bibinfo {author} {\bibfnamefont {M.}~\bibnamefont
			{Warner}},\ }\bibfield  {title} {\bibinfo {title} {Multistage crosslinking of
			nematic networks},\ }\href {https://doi.org/10.1021/MA00116A035} {\bibfield
		{journal} {\bibinfo  {journal} {Macromolecules}\ }\textbf {\bibinfo {volume}
			{28}},\ \bibinfo {pages} {4299} (\bibinfo {year}
		{1995}{\natexlab{b}})}\BibitemShut {NoStop}%
	\bibitem [{\citenamefont {Verwey}\ \emph {et~al.}(1996)\citenamefont {Verwey},
		\citenamefont {Warner},\ and\ \citenamefont {Terentjev}}]{verwey:elastic}%
	\BibitemOpen
	\bibfield  {author} {\bibinfo {author} {\bibfnamefont {G.~C.}\ \bibnamefont
			{Verwey}}, \bibinfo {author} {\bibfnamefont {M.}~\bibnamefont {Warner}},\
		and\ \bibinfo {author} {\bibfnamefont {E.~M.}\ \bibnamefont {Terentjev}},\
	}\bibfield  {title} {\bibinfo {title} {Elastic instability and stripe domains
			in liquid crystalline elastomers},\ }\href
	{https://doi.org/10.1051/jp2:1996130} {\bibfield  {journal} {\bibinfo
			{journal} {J. Phys. II France}\ }\textbf {\bibinfo {volume} {6}},\ \bibinfo
		{pages} {1273} (\bibinfo {year} {1996})}\BibitemShut {NoStop}%
	\bibitem [{\citenamefont {Maier}\ and\ \citenamefont
		{Saupe}(1958)}]{maier:einfache}%
	\BibitemOpen
	\bibfield  {author} {\bibinfo {author} {\bibfnamefont {W.}~\bibnamefont
			{Maier}}\ and\ \bibinfo {author} {\bibfnamefont {A.}~\bibnamefont {Saupe}},\
	}\bibfield  {title} {\bibinfo {title} {Eine einfache molekulare {T}heorie des
			nematischen kristallinfl\"ussigen {Z}ustandes},\ }\href@noop {} {\bibfield
		{journal} {\bibinfo  {journal} {Z. Naturforsch.}\ }\textbf {\bibinfo {volume}
			{13a}},\ \bibinfo {pages} {564} (\bibinfo {year} {1958})},\ \bibinfo {note}
	{translated into English in \cite{sluckin:crystal},
		pp.~381--385.}\BibitemShut {Stop}%
	\bibitem [{\citenamefont {Bai}\ and\ \citenamefont
		{Bhattacharya}(2020)}]{bai:photomechanical}%
	\BibitemOpen
	\bibfield  {author} {\bibinfo {author} {\bibfnamefont {R.}~\bibnamefont
			{Bai}}\ and\ \bibinfo {author} {\bibfnamefont {K.}~\bibnamefont
			{Bhattacharya}},\ }\bibfield  {title} {\bibinfo {title} {Photomechanical
			coupling in photoactive nematic elastomers},\ }\href
	{https://doi.org/https://doi.org/10.1016/j.jmps.2020.104115} {\bibfield
		{journal} {\bibinfo  {journal} {J. Mech. Phys. Solids}\ }\textbf {\bibinfo
			{volume} {144}},\ \bibinfo {pages} {104115} (\bibinfo {year}
		{2020})}\BibitemShut {NoStop}%
	\bibitem [{\citenamefont {Corbett}\ and\ \citenamefont
		{Warner}(2006)}]{corbett:nonlinear}%
	\BibitemOpen
	\bibfield  {author} {\bibinfo {author} {\bibfnamefont {D.}~\bibnamefont
			{Corbett}}\ and\ \bibinfo {author} {\bibfnamefont {M.}~\bibnamefont
			{Warner}},\ }\bibfield  {title} {\bibinfo {title} {Nonlinear photoresponse of
			disordered elastomers},\ }\href
	{https://doi.org/10.1103/PhysRevLett.96.237802} {\bibfield  {journal}
		{\bibinfo  {journal} {Phys. Rev. Lett.}\ }\textbf {\bibinfo {volume} {96}},\
		\bibinfo {pages} {237802} (\bibinfo {year} {2006})}\BibitemShut {NoStop}%
	\bibitem [{\citenamefont {Corbett}\ and\ \citenamefont
		{Warner}(2007)}]{corbett:linear}%
	\BibitemOpen
	\bibfield  {author} {\bibinfo {author} {\bibfnamefont {D.}~\bibnamefont
			{Corbett}}\ and\ \bibinfo {author} {\bibfnamefont {M.}~\bibnamefont
			{Warner}},\ }\bibfield  {title} {\bibinfo {title} {Linear and nonlinear
			photoinduced deformations of cantilevers},\ }\href
	{https://doi.org/10.1103/PhysRevLett.99.174302} {\bibfield  {journal}
		{\bibinfo  {journal} {Phys. Rev. Lett.}\ }\textbf {\bibinfo {volume} {99}},\
		\bibinfo {pages} {174302} (\bibinfo {year} {2007})}\BibitemShut {NoStop}%
	\bibitem [{\citenamefont {Corbett}\ and\ \citenamefont
		{Warner}(2008)}]{corbett:polarization}%
	\BibitemOpen
	\bibfield  {author} {\bibinfo {author} {\bibfnamefont {D.}~\bibnamefont
			{Corbett}}\ and\ \bibinfo {author} {\bibfnamefont {M.}~\bibnamefont
			{Warner}},\ }\bibfield  {title} {\bibinfo {title} {Polarization dependence of
			optically driven polydomain elastomer mechanics},\ }\href
	{https://doi.org/10.1103/PhysRevE.78.061701} {\bibfield  {journal} {\bibinfo
			{journal} {Phys. Rev. E}\ }\textbf {\bibinfo {volume} {78}},\ \bibinfo
		{pages} {061701} (\bibinfo {year} {2008})}\BibitemShut {NoStop}%
	\bibitem [{\citenamefont {Mahimwalla}\ \emph {et~al.}(2012)\citenamefont
		{Mahimwalla}, \citenamefont {Yager}, \citenamefont {ichi Mamiya},
		\citenamefont {Shishido}, \citenamefont {Priimagi},\ and\ \citenamefont
		{Barrett}}]{mahimwalla:azobenzene}%
	\BibitemOpen
	\bibfield  {author} {\bibinfo {author} {\bibfnamefont {Z.}~\bibnamefont
			{Mahimwalla}}, \bibinfo {author} {\bibfnamefont {K.~G.}\ \bibnamefont
			{Yager}}, \bibinfo {author} {\bibfnamefont {J.}~\bibnamefont {ichi Mamiya}},
		\bibinfo {author} {\bibfnamefont {A.}~\bibnamefont {Shishido}}, \bibinfo
		{author} {\bibfnamefont {A.}~\bibnamefont {Priimagi}},\ and\ \bibinfo
		{author} {\bibfnamefont {C.~J.}\ \bibnamefont {Barrett}},\ }\bibfield
	{title} {\bibinfo {title} {Azobenzene photomechanics: prospects and potential
			applications},\ }\href
	{https://doi.org/https://doi.org/10.1007/s00289-012-0792-0} {\bibfield
		{journal} {\bibinfo  {journal} {Polym. Bull.}\ }\textbf {\bibinfo {volume}
			{69}},\ \bibinfo {pages} {967} (\bibinfo {year} {2012})}\BibitemShut
	{NoStop}%
	\bibitem [{\citenamefont {Ube}\ and\ \citenamefont
		{Ikeda}(2014)}]{ube:photomobile}%
	\BibitemOpen
	\bibfield  {author} {\bibinfo {author} {\bibfnamefont {T.}~\bibnamefont
			{Ube}}\ and\ \bibinfo {author} {\bibfnamefont {T.}~\bibnamefont {Ikeda}},\
	}\bibfield  {title} {\bibinfo {title} {Photomobile polymer materials with
			crosslinked liquid-crystalline structures: Molecular design, fabrication, and
			functions},\ }\href {https://doi.org/https://doi.org/10.1002/anie.201400513}
	{\bibfield  {journal} {\bibinfo  {journal} {Angew. Chem. Int. Ed.}\ }\textbf
		{\bibinfo {volume} {53}},\ \bibinfo {pages} {10290} (\bibinfo {year}
		{2014})}\BibitemShut {NoStop}%
	\bibitem [{\citenamefont {White}(2018)}]{white:photomechanical}%
	\BibitemOpen
	\bibfield  {author} {\bibinfo {author} {\bibfnamefont {T.~J.}\ \bibnamefont
			{White}},\ }\bibfield  {title} {\bibinfo {title} {Photomechanical effects in
			liquid crystalline polymer networks and elastomers},\ }\href
	{https://doi.org/https://doi.org/10.1002/polb.24576} {\bibfield  {journal}
		{\bibinfo  {journal} {J. Polym. Sci. Part B: Polym. Phys.}\ }\textbf
		{\bibinfo {volume} {56}},\ \bibinfo {pages} {695} (\bibinfo {year}
		{2018})}\BibitemShut {NoStop}%
	\bibitem [{\citenamefont {Ula}\ \emph {et~al.}(2018)\citenamefont {Ula},
		\citenamefont {Traugutt}, \citenamefont {Volpe}, \citenamefont {Patel},
		\citenamefont {Yu},\ and\ \citenamefont {Yakacki}}]{ula:liquid}%
	\BibitemOpen
	\bibfield  {author} {\bibinfo {author} {\bibfnamefont {S.~W.}\ \bibnamefont
			{Ula}}, \bibinfo {author} {\bibfnamefont {N.~A.}\ \bibnamefont {Traugutt}},
		\bibinfo {author} {\bibfnamefont {R.~H.}\ \bibnamefont {Volpe}}, \bibinfo
		{author} {\bibfnamefont {R.~R.}\ \bibnamefont {Patel}}, \bibinfo {author}
		{\bibfnamefont {K.}~\bibnamefont {Yu}},\ and\ \bibinfo {author}
		{\bibfnamefont {C.~M.}\ \bibnamefont {Yakacki}},\ }\bibfield  {title}
	{\bibinfo {title} {Liquid crystal elastomers: an introduction and review of
			emerging technologies},\ }\href
	{https://doi.org/10.1080/21680396.2018.1530155} {\bibfield  {journal}
		{\bibinfo  {journal} {Liquid Cryst. Rev.}\ }\textbf {\bibinfo {volume} {6}},\
		\bibinfo {pages} {78} (\bibinfo {year} {2018})}\BibitemShut {NoStop}%
	\bibitem [{\citenamefont {Pang}\ \emph {et~al.}(2019)\citenamefont {Pang},
		\citenamefont {Lv}, \citenamefont {Zhu}, \citenamefont {Qin},\ and\
		\citenamefont {Yu}}]{pang:photodeformable}%
	\BibitemOpen
	\bibfield  {author} {\bibinfo {author} {\bibfnamefont {X.}~\bibnamefont
			{Pang}}, \bibinfo {author} {\bibfnamefont {J.-a.}\ \bibnamefont {Lv}},
		\bibinfo {author} {\bibfnamefont {C.}~\bibnamefont {Zhu}}, \bibinfo {author}
		{\bibfnamefont {L.}~\bibnamefont {Qin}},\ and\ \bibinfo {author}
		{\bibfnamefont {Y.}~\bibnamefont {Yu}},\ }\bibfield  {title} {\bibinfo
		{title} {Photodeformable azobenzene-containing real polymers and soft
			actuators},\ }\href {https://doi.org/https://doi.org/10.1002/adma.201904224}
	{\bibfield  {journal} {\bibinfo  {journal} {Adv. Mater.}\ }\textbf {\bibinfo
			{volume} {31}},\ \bibinfo {pages} {1904224} (\bibinfo {year}
		{2019})}\BibitemShut {NoStop}%
	\bibitem [{\citenamefont {Kuenstler}\ and\ \citenamefont
		{Hayward}(2019)}]{kuenstler:light}%
	\BibitemOpen
	\bibfield  {author} {\bibinfo {author} {\bibfnamefont {A.~S.}\ \bibnamefont
			{Kuenstler}}\ and\ \bibinfo {author} {\bibfnamefont {R.~C.}\ \bibnamefont
			{Hayward}},\ }\bibfield  {title} {\bibinfo {title} {Light-induced shape
			morphing of thin films},\ }\href
	{https://doi.org/https://doi.org/10.1016/j.cocis.2019.01.009} {\bibfield
		{journal} {\bibinfo  {journal} {Curr. Opin. Colloid \& Interface Sci.}\
		}\textbf {\bibinfo {volume} {40}},\ \bibinfo {pages} {70} (\bibinfo {year}
		{2019})}\BibitemShut {NoStop}%
	\bibitem [{\citenamefont {Warner}(2020)}]{warner:topographic}%
	\BibitemOpen
	\bibfield  {author} {\bibinfo {author} {\bibfnamefont {M.}~\bibnamefont
			{Warner}},\ }\bibfield  {title} {\bibinfo {title} {Topographic mechanics and
			applications of liquid crystalline solids},\ }\href
	{https://doi.org/10.1146/annurev-conmatphys-031119-050738} {\bibfield
		{journal} {\bibinfo  {journal} {Annu. Rev. Condens. Matter Phys.}\ }\textbf
		{\bibinfo {volume} {11}},\ \bibinfo {pages} {125} (\bibinfo {year}
		{2020})}\BibitemShut {NoStop}%
	\bibitem [{\citenamefont {Anderson}\ \emph {et~al.}(1999)\citenamefont
		{Anderson}, \citenamefont {Carlson},\ and\ \citenamefont
		{Fried}}]{anderson:continuum}%
	\BibitemOpen
	\bibfield  {author} {\bibinfo {author} {\bibfnamefont {D.~R.}\ \bibnamefont
			{Anderson}}, \bibinfo {author} {\bibfnamefont {D.~E.}\ \bibnamefont
			{Carlson}},\ and\ \bibinfo {author} {\bibfnamefont {E.}~\bibnamefont
			{Fried}},\ }\bibfield  {title} {\bibinfo {title} {A continuum-mechanical
			theory for nematic elastomers},\ }\href
	{https://doi.org/10.1023/A:1007647913363} {\bibfield  {journal} {\bibinfo
			{journal} {J. Elast.}\ }\textbf {\bibinfo {volume} {56}},\ \bibinfo {pages}
		{33} (\bibinfo {year} {1999})}\BibitemShut {NoStop}%
	\bibitem [{\citenamefont {Zhang}\ \emph {et~al.}(2019)\citenamefont {Zhang},
		\citenamefont {Xuan}, \citenamefont {Jiang},\ and\ \citenamefont
		{Huo}}]{zhang:continuum}%
	\BibitemOpen
	\bibfield  {author} {\bibinfo {author} {\bibfnamefont {Y.}~\bibnamefont
			{Zhang}}, \bibinfo {author} {\bibfnamefont {C.}~\bibnamefont {Xuan}},
		\bibinfo {author} {\bibfnamefont {Y.}~\bibnamefont {Jiang}},\ and\ \bibinfo
		{author} {\bibfnamefont {Y.}~\bibnamefont {Huo}},\ }\bibfield  {title}
	{\bibinfo {title} {Continuum mechanical modeling of liquid crystal elastomers
			as dissipative ordered solids},\ }\href
	{https://doi.org/10.1016/j.jmps.2019.02.018} {\bibfield  {journal} {\bibinfo
			{journal} {J. Mech. Phys. Solids}\ }\textbf {\bibinfo {volume} {126}},\
		\bibinfo {pages} {285} (\bibinfo {year} {2019})}\BibitemShut {NoStop}%
	\bibitem [{\citenamefont {Mihai}\ \emph {et~al.}(2021)\citenamefont {Mihai},
		\citenamefont {Wang}, \citenamefont {Guilleminot},\ and\ \citenamefont
		{Goriely}}]{mihai:nematic}%
	\BibitemOpen
	\bibfield  {author} {\bibinfo {author} {\bibfnamefont {L.~A.}\ \bibnamefont
			{Mihai}}, \bibinfo {author} {\bibfnamefont {H.}~\bibnamefont {Wang}},
		\bibinfo {author} {\bibfnamefont {J.}~\bibnamefont {Guilleminot}},\ and\
		\bibinfo {author} {\bibfnamefont {A.}~\bibnamefont {Goriely}},\ }\bibfield
	{title} {\bibinfo {title} {Nematic liquid crystalline elastomers are
			aeolotropic materials},\ }\href {https://doi.org/10.1098/rspa.2021.0259}
	{\bibfield  {journal} {\bibinfo  {journal} {Proc. R. Soc. London A}\ }\textbf
		{\bibinfo {volume} {477}},\ \bibinfo {pages} {20210259} (\bibinfo {year}
		{2021})}\BibitemShut {NoStop}%
	\bibitem [{\citenamefont {Korley}\ and\ \citenamefont
		{Ware}(2021)}]{korley:introduction}%
	\BibitemOpen
	\bibfield  {author} {\bibinfo {author} {\bibfnamefont {L.~T.~J.}\
			\bibnamefont {Korley}}\ and\ \bibinfo {author} {\bibfnamefont {T.~H.}\
			\bibnamefont {Ware}},\ }\bibfield  {title} {\bibinfo {title} {Introduction to
			special topic: {P}rogrammable liquid crystal elastomers},\ }\href
	{https://doi.org/10.1063/5.0078455} {\bibfield  {journal} {\bibinfo
			{journal} {J. Appl. Phys.}\ }\textbf {\bibinfo {volume} {130}},\ \bibinfo
		{pages} {220401} (\bibinfo {year} {2021})}\BibitemShut {NoStop}%
	\bibitem [{\citenamefont {White}\ and\ \citenamefont
		{Broer}(2015)}]{white:programmable}%
	\BibitemOpen
	\bibfield  {author} {\bibinfo {author} {\bibfnamefont {T.~J.}\ \bibnamefont
			{White}}\ and\ \bibinfo {author} {\bibfnamefont {D.~J.}\ \bibnamefont
			{Broer}},\ }\bibfield  {title} {\bibinfo {title} {Programmable and adaptive
			mechanics with liquid crystal polymer networks and elastomers},\ }\href
	{https://doi.org/https://doi.org/10.1038/nmat4433} {\bibfield  {journal}
		{\bibinfo  {journal} {Nature Mater.}\ }\textbf {\bibinfo {volume} {14}},\
		\bibinfo {pages} {1087} (\bibinfo {year} {2015})}\BibitemShut {NoStop}%
	\bibitem [{\citenamefont {Modes}\ \emph {et~al.}(2010)\citenamefont {Modes},
		\citenamefont {Bhattacharya},\ and\ \citenamefont
		{Warner}}]{modes:disclination}%
	\BibitemOpen
	\bibfield  {author} {\bibinfo {author} {\bibfnamefont {C.~D.}\ \bibnamefont
			{Modes}}, \bibinfo {author} {\bibfnamefont {K.}~\bibnamefont
			{Bhattacharya}},\ and\ \bibinfo {author} {\bibfnamefont {M.}~\bibnamefont
			{Warner}},\ }\bibfield  {title} {\bibinfo {title} {Disclination-mediated
			thermo-optical response in nematic glass sheets},\ }\href
	{https://doi.org/10.1103/PhysRevE.81.060701} {\bibfield  {journal} {\bibinfo
			{journal} {Phys. Rev. E}\ }\textbf {\bibinfo {volume} {81}},\ \bibinfo
		{pages} {060701} (\bibinfo {year} {2010})}\BibitemShut {NoStop}%
	\bibitem [{\citenamefont {Modes}\ and\ \citenamefont
		{Warner}(2011)}]{modes:blueprinting}%
	\BibitemOpen
	\bibfield  {author} {\bibinfo {author} {\bibfnamefont {C.~D.}\ \bibnamefont
			{Modes}}\ and\ \bibinfo {author} {\bibfnamefont {M.}~\bibnamefont {Warner}},\
	}\bibfield  {title} {\bibinfo {title} {Blueprinting nematic glass:
			{S}ystematically constructing and combining active points of curvature for
			emergent morphology},\ }\href {https://doi.org/10.1103/PhysRevE.84.021711}
	{\bibfield  {journal} {\bibinfo  {journal} {Phys. Rev. E}\ }\textbf {\bibinfo
			{volume} {84}},\ \bibinfo {pages} {021711} (\bibinfo {year}
		{2011})}\BibitemShut {NoStop}%
	\bibitem [{\citenamefont {Kutter}\ and\ \citenamefont
		{Terentjev}(2001)}]{kutter:tube}%
	\BibitemOpen
	\bibfield  {author} {\bibinfo {author} {\bibfnamefont {S.}~\bibnamefont
			{Kutter}}\ and\ \bibinfo {author} {\bibfnamefont {E.}~\bibnamefont
			{Terentjev}},\ }\bibfield  {title} {\bibinfo {title} {Tube model for the
			elasticity of entangled nematic rubbers},\ }\href
	{https://doi.org/10.1007/s101890170004} {\bibfield  {journal} {\bibinfo
			{journal} {Eur. Phys. J. E}\ }\textbf {\bibinfo {volume} {6}},\ \bibinfo
		{pages} {221} (\bibinfo {year} {2001})}\BibitemShut {NoStop}%
	\bibitem [{\citenamefont {Edwards}(1977)}]{edwards:theory}%
	\BibitemOpen
	\bibfield  {author} {\bibinfo {author} {\bibfnamefont {S.~F.}\ \bibnamefont
			{Edwards}},\ }\bibfield  {title} {\bibinfo {title} {The theory of rubber
			elasticity},\ }\href {https://doi.org/https://doi.org/10.1002/pi.4980090209}
	{\bibfield  {journal} {\bibinfo  {journal} {Brit. Polym. J.}\ }\textbf
		{\bibinfo {volume} {9}},\ \bibinfo {pages} {140} (\bibinfo {year}
		{1977})}\BibitemShut {NoStop}%
	\bibitem [{\citenamefont {Ozenda}\ \emph {et~al.}(2020)\citenamefont {Ozenda},
		\citenamefont {Sonnet},\ and\ \citenamefont {Virga}}]{ozenda:blend}%
	\BibitemOpen
	\bibfield  {author} {\bibinfo {author} {\bibfnamefont {O.}~\bibnamefont
			{Ozenda}}, \bibinfo {author} {\bibfnamefont {A.~M.}\ \bibnamefont {Sonnet}},\
		and\ \bibinfo {author} {\bibfnamefont {E.~G.}\ \bibnamefont {Virga}},\
	}\bibfield  {title} {\bibinfo {title} {A blend of stretching and bending in
			nematic polymer networks},\ }\href
	{https://doi.org/https://doi.org/10.1039/D0SM00642D} {\bibfield  {journal}
		{\bibinfo  {journal} {Soft Matter}\ }\textbf {\bibinfo {volume} {16}},\
		\bibinfo {pages} {8877} (\bibinfo {year} {2020})}\BibitemShut {NoStop}%
	\bibitem [{\citenamefont {Ozenda}\ and\ \citenamefont
		{Virga}(2021)}]{ozenda:kirchhoff}%
	\BibitemOpen
	\bibfield  {author} {\bibinfo {author} {\bibfnamefont {O.}~\bibnamefont
			{Ozenda}}\ and\ \bibinfo {author} {\bibfnamefont {E.~G.}\ \bibnamefont
			{Virga}},\ }\bibfield  {title} {\bibinfo {title} {On the {K}irchhoff-{L}ove
			hypothesis (revised and vindicated)},\ }\href
	{https://doi.org/https://doi.org/10.1007/s10659-021-09819-7} {\bibfield
		{journal} {\bibinfo  {journal} {J. Elast.}\ }\textbf {\bibinfo {volume}
			{143}},\ \bibinfo {pages} {359} (\bibinfo {year} {2021})}\BibitemShut
	{NoStop}%
	\bibitem [{\citenamefont {Stoker}(1969)}]{stoker:differential}%
	\BibitemOpen
	\bibfield  {author} {\bibinfo {author} {\bibfnamefont {J.~J.}\ \bibnamefont
			{Stoker}},\ }\href@noop {} {\emph {\bibinfo {title} {Differential
				Geometry}}},\ \bibinfo {series} {Pure and Applied Mathematics}, Vol.~\bibinfo
	{volume} {XX}\ (\bibinfo  {publisher} {Wiley-Interscience},\ \bibinfo
	{address} {New York},\ \bibinfo {year} {1969})\BibitemShut {NoStop}%
	\bibitem [{\citenamefont {Pedrini}\ and\ \citenamefont
		{Virga}(2021{\natexlab{a}})}]{pedrini:ridge}%
	\BibitemOpen
	\bibfield  {author} {\bibinfo {author} {\bibfnamefont {A.}~\bibnamefont
			{Pedrini}}\ and\ \bibinfo {author} {\bibfnamefont {E.~G.}\ \bibnamefont
			{Virga}},\ }\bibfield  {title} {\bibinfo {title} {Ridge energy for thin
			nematic polymer networks},\ }\href
	{https://doi.org/https://doi.org/10.1140/epje/s10189-021-00012-1} {\bibfield
		{journal} {\bibinfo  {journal} {Eur. Phys. J. E}\ }\textbf {\bibinfo {volume}
			{44}},\ \bibinfo {pages} {7} (\bibinfo {year}
		{2021}{\natexlab{a}})}\BibitemShut {NoStop}%
	\bibitem [{\citenamefont {Pedrini}\ and\ \citenamefont
		{Virga}(2021{\natexlab{b}})}]{pedrini:ridge_JPA}%
	\BibitemOpen
	\bibfield  {author} {\bibinfo {author} {\bibfnamefont {A.}~\bibnamefont
			{Pedrini}}\ and\ \bibinfo {author} {\bibfnamefont {E.~G.}\ \bibnamefont
			{Virga}},\ }\bibfield  {title} {\bibinfo {title} {Ridge approximation for
			thin nematic polymer networks},\ }\href {https://doi.org/10.1063/5.0045070}
	{\bibfield  {journal} {\bibinfo  {journal} {J. Appl. Phys.}\ }\textbf
		{\bibinfo {volume} {129}},\ \bibinfo {pages} {184701} (\bibinfo {year}
		{2021}{\natexlab{b}})}\BibitemShut {NoStop}%
	\bibitem [{\citenamefont {Audoly}\ and\ \citenamefont
		{Pomeau}(2010)}]{audoly:elasticity}%
	\BibitemOpen
	\bibfield  {author} {\bibinfo {author} {\bibfnamefont {B.}~\bibnamefont
			{Audoly}}\ and\ \bibinfo {author} {\bibfnamefont {Y.}~\bibnamefont
			{Pomeau}},\ }\href@noop {} {\emph {\bibinfo {title} {Elasticity and
				Geometry}}}\ (\bibinfo  {publisher} {Oxford University Press},\ \bibinfo
	{address} {Oxford},\ \bibinfo {year} {2010})\BibitemShut {NoStop}%
	\bibitem [{\citenamefont {Grossman}\ \emph {et~al.}(2016)\citenamefont
		{Grossman}, \citenamefont {Sharon},\ and\ \citenamefont
		{Diamant}}]{grossman:elasticity}%
	\BibitemOpen
	\bibfield  {author} {\bibinfo {author} {\bibfnamefont {D.}~\bibnamefont
			{Grossman}}, \bibinfo {author} {\bibfnamefont {E.}~\bibnamefont {Sharon}},\
		and\ \bibinfo {author} {\bibfnamefont {H.}~\bibnamefont {Diamant}},\
	}\bibfield  {title} {\bibinfo {title} {Elasticity and fluctuations of
			frustrated nanoribbons},\ }\href
	{https://doi.org/10.1103/PhysRevLett.116.258105} {\bibfield  {journal}
		{\bibinfo  {journal} {Phys. Rev. Lett.}\ }\textbf {\bibinfo {volume} {116}},\
		\bibinfo {pages} {258105} (\bibinfo {year} {2016})}\BibitemShut {NoStop}%
	\bibitem [{\citenamefont {Agostiniani}\ \emph {et~al.}(2017)\citenamefont
		{Agostiniani}, \citenamefont {De{S}imone},\ and\ \citenamefont
		{Koumatos}}]{agostiniani:shape}%
	\BibitemOpen
	\bibfield  {author} {\bibinfo {author} {\bibfnamefont {V.}~\bibnamefont
			{Agostiniani}}, \bibinfo {author} {\bibfnamefont {A.}~\bibnamefont
			{De{S}imone}},\ and\ \bibinfo {author} {\bibfnamefont {K.}~\bibnamefont
			{Koumatos}},\ }\bibfield  {title} {\bibinfo {title} {Shape programming for
			narrow ribbons of nematic elastomers},\ }\href
	{https://doi.org/https://doi.org/10.1007/s10659-016-9594-1} {\bibfield
		{journal} {\bibinfo  {journal} {J. Elast.}\ }\textbf {\bibinfo {volume}
			{127}},\ \bibinfo {pages} {1} (\bibinfo {year} {2017})}\BibitemShut {NoStop}%
	\bibitem [{\citenamefont {Agostiniani}\ and\ \citenamefont
		{De{S}imone}(2020)}]{agostiniani:rigorous}%
	\BibitemOpen
	\bibfield  {author} {\bibinfo {author} {\bibfnamefont {V.}~\bibnamefont
			{Agostiniani}}\ and\ \bibinfo {author} {\bibfnamefont {A.}~\bibnamefont
			{De{S}imone}},\ }\bibfield  {title} {\bibinfo {title} {Rigorous derivation of
			active plate models for thin sheets of nematic elastomers},\ }\href
	{https://doi.org/https://doi.org/10.1177/1081286517699991} {\bibfield
		{journal} {\bibinfo  {journal} {Math. Mech. Solids}\ }\textbf {\bibinfo
			{volume} {25}},\ \bibinfo {pages} {1804} (\bibinfo {year}
		{2020})}\BibitemShut {NoStop}%
	\bibitem [{\citenamefont {Efrati}\ \emph {et~al.}(2009)\citenamefont {Efrati},
		\citenamefont {Sharon},\ and\ \citenamefont {Kupferman}}]{efrati:elastic}%
	\BibitemOpen
	\bibfield  {author} {\bibinfo {author} {\bibfnamefont {E.}~\bibnamefont
			{Efrati}}, \bibinfo {author} {\bibfnamefont {E.}~\bibnamefont {Sharon}},\
		and\ \bibinfo {author} {\bibfnamefont {R.}~\bibnamefont {Kupferman}},\
	}\bibfield  {title} {\bibinfo {title} {Elastic theory of unconstrained
			non-{E}uclidean plates},\ }\href
	{https://doi.org/https://doi.org/10.1016/j.jmps.2008.12.004} {\bibfield
		{journal} {\bibinfo  {journal} {J. Mech. Phys. Solids}\ }\textbf {\bibinfo
			{volume} {57}},\ \bibinfo {pages} {762} (\bibinfo {year} {2009})}\BibitemShut
	{NoStop}%
	\bibitem [{\citenamefont {Mihai}\ and\ \citenamefont
		{Goriely}(2020)}]{mihai:plate}%
	\BibitemOpen
	\bibfield  {author} {\bibinfo {author} {\bibfnamefont {L.~A.}\ \bibnamefont
			{Mihai}}\ and\ \bibinfo {author} {\bibfnamefont {A.}~\bibnamefont
			{Goriely}},\ }\bibfield  {title} {\bibinfo {title} {A plate theory for
			nematic liquid crystalline solids},\ }\href
	{https://doi.org/https://doi.org/10.1016/j.jmps.2020.104101} {\bibfield
		{journal} {\bibinfo  {journal} {J. Mech. Phys. Solids}\ }\textbf {\bibinfo
			{volume} {144}},\ \bibinfo {pages} {104101} (\bibinfo {year}
		{2020})}\BibitemShut {NoStop}%
	\bibitem [{\citenamefont {Warner}\ \emph {et~al.}(1988)\citenamefont {Warner},
		\citenamefont {Gelling},\ and\ \citenamefont {Vilgis}}]{warner:theory}%
	\BibitemOpen
	\bibfield  {author} {\bibinfo {author} {\bibfnamefont {M.}~\bibnamefont
			{Warner}}, \bibinfo {author} {\bibfnamefont {K.~P.}\ \bibnamefont
			{Gelling}},\ and\ \bibinfo {author} {\bibfnamefont {T.~A.}\ \bibnamefont
			{Vilgis}},\ }\bibfield  {title} {\bibinfo {title} {Theory of nematic
			networks},\ }\href {https://doi.org/10.1063/1.453852} {\bibfield  {journal}
		{\bibinfo  {journal} {J. Chem. Phys.}\ }\textbf {\bibinfo {volume} {88}},\
		\bibinfo {pages} {4008} (\bibinfo {year} {1988})}\BibitemShut {NoStop}%
	\bibitem [{\citenamefont {Warner}\ and\ \citenamefont
		{Wang}(1991)}]{warner:elasticity}%
	\BibitemOpen
	\bibfield  {author} {\bibinfo {author} {\bibfnamefont {M.}~\bibnamefont
			{Warner}}\ and\ \bibinfo {author} {\bibfnamefont {X.~J.}\ \bibnamefont
			{Wang}},\ }\bibfield  {title} {\bibinfo {title} {Elasticity and phase
			behavior of nematic elastomers},\ }\href
	{https://doi.org/10.1021/ma00017a033} {\bibfield  {journal} {\bibinfo
			{journal} {Macromolecules}\ }\textbf {\bibinfo {volume} {24}},\ \bibinfo
		{pages} {4932} (\bibinfo {year} {1991})}\BibitemShut {NoStop}%
	\bibitem [{\citenamefont {Warner}\ and\ \citenamefont
		{Terentjev}(1996)}]{warner:nematic_elastomer}%
	\BibitemOpen
	\bibfield  {author} {\bibinfo {author} {\bibfnamefont {M.}~\bibnamefont
			{Warner}}\ and\ \bibinfo {author} {\bibfnamefont {E.}~\bibnamefont
			{Terentjev}},\ }\bibfield  {title} {\bibinfo {title} {Nematic
			elastomers---{A} new state of matter?},\ }\href
	{https://doi.org/https://doi.org/10.1016/S0079-6700(96)00013-5} {\bibfield
		{journal} {\bibinfo  {journal} {Prog. Polym. Sci.}\ }\textbf {\bibinfo
			{volume} {21}},\ \bibinfo {pages} {853} (\bibinfo {year} {1996})}\BibitemShut
	{NoStop}%
	\bibitem [{\citenamefont {Treloar}(2005)}]{treloar:non-linear_third}%
	\BibitemOpen
	\bibfield  {author} {\bibinfo {author} {\bibfnamefont {L.~R.~G.}\
			\bibnamefont {Treloar}},\ }\href@noop {} {\emph {\bibinfo {title} {The
				Physics of Rubber Elasticity}}},\ \bibinfo {edition} {3rd}\ ed.,\ Oxford
	Classic Texts in the Physical Sciences\ (\bibinfo  {publisher} {Oxford
		University Press},\ \bibinfo {address} {Oxford},\ \bibinfo {year}
	{2005})\BibitemShut {NoStop}%
	\bibitem [{\citenamefont {Nguyen}\ and\ \citenamefont
		{Selinger}(2017)}]{nguyen:theory}%
	\BibitemOpen
	\bibfield  {author} {\bibinfo {author} {\bibfnamefont {T.-S.}\ \bibnamefont
			{Nguyen}}\ and\ \bibinfo {author} {\bibfnamefont {J.}~\bibnamefont
			{Selinger}},\ }\bibfield  {title} {\bibinfo {title} {Theory of liquid crystal
			elastomers and polymer networks},\ }\href
	{https://doi.org/10.1140/epje/i2017-11569-5} {\bibfield  {journal} {\bibinfo
			{journal} {Eur. Phys. J. E}\ }\textbf {\bibinfo {volume} {40}},\ \bibinfo
		{pages} {76} (\bibinfo {year} {2017})}\BibitemShut {NoStop}%
	\bibitem [{\citenamefont {Dias}\ and\ \citenamefont
		{Audoly}(2015)}]{dias:wunderlich}%
	\BibitemOpen
	\bibfield  {author} {\bibinfo {author} {\bibfnamefont {M.~A.}\ \bibnamefont
			{Dias}}\ and\ \bibinfo {author} {\bibfnamefont {B.}~\bibnamefont {Audoly}},\
	}\bibfield  {title} {\bibinfo {title} {``{W}underlich, meet {K}irchhoff'':
			{A} general and unified description of elastic ribbons and thin rods},\
	}\href {https://doi.org/https://doi.org/10.1007/s10659-014-9487-0} {\bibfield
		{journal} {\bibinfo  {journal} {J. Elast.}\ }\textbf {\bibinfo {volume}
			{119}},\ \bibinfo {pages} {49} (\bibinfo {year} {2015})}\BibitemShut
	{NoStop}%
	\bibitem [{\citenamefont {Chen}\ and\ \citenamefont
		{Fried}(2016)}]{chen:moebius}%
	\BibitemOpen
	\bibfield  {author} {\bibinfo {author} {\bibfnamefont {Y.-C.}\ \bibnamefont
			{Chen}}\ and\ \bibinfo {author} {\bibfnamefont {E.}~\bibnamefont {Fried}},\
	}\bibfield  {title} {\bibinfo {title} {M\"{o}bius bands, unstretchable
			material sheets and developable surfaces},\ }\href
	{https://doi.org/10.1098/rspa.2016.0459} {\bibfield  {journal} {\bibinfo
			{journal} {Proc. R. Soc. London A}\ }\textbf {\bibinfo {volume} {472}},\
		\bibinfo {pages} {20160459} (\bibinfo {year} {2016})}\BibitemShut {NoStop}%
	\bibitem [{\citenamefont {Chen}\ \emph
		{et~al.}(2018{\natexlab{a}})\citenamefont {Chen}, \citenamefont {Fosdick},\
		and\ \citenamefont {Fried}}]{chen:issues}%
	\BibitemOpen
	\bibfield  {author} {\bibinfo {author} {\bibfnamefont {Y.-C.}\ \bibnamefont
			{Chen}}, \bibinfo {author} {\bibfnamefont {R.}~\bibnamefont {Fosdick}},\ and\
		\bibinfo {author} {\bibfnamefont {E.}~\bibnamefont {Fried}},\ }\bibfield
	{title} {\bibinfo {title} {Issues concerning isometric deformations of planar
			regions to curved surfaces},\ }\href
	{https://doi.org/https://doi.org/10.1007/s10659-017-9662-1} {\bibfield
		{journal} {\bibinfo  {journal} {J. Elast.}\ }\textbf {\bibinfo {volume}
			{132}},\ \bibinfo {pages} {1} (\bibinfo {year}
		{2018}{\natexlab{a}})}\BibitemShut {NoStop}%
	\bibitem [{\citenamefont {van~der Heijden}\ and\ \citenamefont
		{Starostin}(2022)}]{heijden:comment}%
	\BibitemOpen
	\bibfield  {author} {\bibinfo {author} {\bibfnamefont {G.~H.~M.}\
			\bibnamefont {van~der Heijden}}\ and\ \bibinfo {author} {\bibfnamefont
			{E.~L.}\ \bibnamefont {Starostin}},\ }\bibfield  {title} {\bibinfo {title}
		{Comment on {Y.-C. C}hen, {E. F}ried, {M}\"obius bands, unstretchable
			material sheets and developable surfaces. \emph{{P}roc. {R}. {S}oc. {A}}
			\textbf{472}, 20160459 (2016)},\ }\href
	{https://doi.org/10.1098/rspa.2021.0629} {\bibfield  {journal} {\bibinfo
			{journal} {Proc. R. Soc. A}\ }\textbf {\bibinfo {volume} {478}},\ \bibinfo
		{pages} {20210629} (\bibinfo {year} {2022})}\BibitemShut {NoStop}%
	\bibitem [{\citenamefont {Chen}\ \emph {et~al.}(2022)\citenamefont {Chen},
		\citenamefont {Fosdick},\ and\ \citenamefont {Fried}}]{chen:reply}%
	\BibitemOpen
	\bibfield  {author} {\bibinfo {author} {\bibfnamefont {Y.-C.}\ \bibnamefont
			{Chen}}, \bibinfo {author} {\bibfnamefont {R.}~\bibnamefont {Fosdick}},\ and\
		\bibinfo {author} {\bibfnamefont {E.}~\bibnamefont {Fried}},\ }\bibfield
	{title} {\bibinfo {title} {Reply to the comment of van der {H}eijden and
			{S}tarostin},\ }\href {https://doi.org/10.1098/rspa.2021.0856} {\bibfield
		{journal} {\bibinfo  {journal} {Proc. R. Soc. London A}\ }\textbf {\bibinfo
			{volume} {478}},\ \bibinfo {pages} {20210856} (\bibinfo {year}
		{2022})}\BibitemShut {NoStop}%
	\bibitem [{\citenamefont {Chen}\ \emph {et~al.}(2015)\citenamefont {Chen},
		\citenamefont {Fosdick},\ and\ \citenamefont
		{Fried}}]{chen:representation_2015}%
	\BibitemOpen
	\bibfield  {author} {\bibinfo {author} {\bibfnamefont {Y.-C.}\ \bibnamefont
			{Chen}}, \bibinfo {author} {\bibfnamefont {R.}~\bibnamefont {Fosdick}},\ and\
		\bibinfo {author} {\bibfnamefont {E.}~\bibnamefont {Fried}},\ }\bibfield
	{title} {\bibinfo {title} {Representation for a smooth isometric mapping from
			a connected planar domain to a surface},\ }\href
	{https://doi.org/https://doi.org/10.1007/s10659-015-9521-x} {\bibfield
		{journal} {\bibinfo  {journal} {J. Elast.}\ }\textbf {\bibinfo {volume}
			{119}},\ \bibinfo {pages} {335} (\bibinfo {year} {2015})}\BibitemShut
	{NoStop}%
	\bibitem [{\citenamefont {Schwarz}(1990)}]{schwarz:dark}%
	\BibitemOpen
	\bibfield  {author} {\bibinfo {author} {\bibfnamefont {G.~E.}\ \bibnamefont
			{Schwarz}},\ }\bibfield  {title} {\bibinfo {title} {The dark side of the
			{M}oebius strip},\ }\href {https://doi.org/10.2307/2324325} {\bibfield
		{journal} {\bibinfo  {journal} {Am. Math. Mon.}\ }\textbf {\bibinfo {volume}
			{97}},\ \bibinfo {pages} {890} (\bibinfo {year} {1990})}\BibitemShut
	{NoStop}%
	\bibitem [{\citenamefont {Mahadevan}\ and\ \citenamefont
		{Keller}(1993)}]{mahadevan:shape}%
	\BibitemOpen
	\bibfield  {author} {\bibinfo {author} {\bibfnamefont {L.}~\bibnamefont
			{Mahadevan}}\ and\ \bibinfo {author} {\bibfnamefont {J.~B.}\ \bibnamefont
			{Keller}},\ }\bibfield  {title} {\bibinfo {title} {The shape of a
			{M}{\"o}bius band},\ }\href {https://doi.org/10.1098/rspa.1993.0009}
	{\bibfield  {journal} {\bibinfo  {journal} {Proc. R. Soc. London A}\ }\textbf
		{\bibinfo {volume} {440}},\ \bibinfo {pages} {149} (\bibinfo {year}
		{1993})}\BibitemShut {NoStop}%
	\bibitem [{\citenamefont {Randrup}\ and\ \citenamefont
		{R{\o}gen}(1996)}]{randrup:sides}%
	\BibitemOpen
	\bibfield  {author} {\bibinfo {author} {\bibfnamefont {T.}~\bibnamefont
			{Randrup}}\ and\ \bibinfo {author} {\bibfnamefont {P.}~\bibnamefont
			{R{\o}gen}},\ }\bibfield  {title} {\bibinfo {title} {Sides of the {M}\"obius
			strip},\ }\href {https://doi.org/https://doi.org/10.1007/BF01268871}
	{\bibfield  {journal} {\bibinfo  {journal} {Arch. Math.}\ }\textbf {\bibinfo
			{volume} {66}},\ \bibinfo {pages} {511} (\bibinfo {year} {1996})}\BibitemShut
	{NoStop}%
	\bibitem [{\citenamefont {Hangan}(2005)}]{hangan:elastic}%
	\BibitemOpen
	\bibfield  {author} {\bibinfo {author} {\bibfnamefont {T.}~\bibnamefont
			{Hangan}},\ }\bibfield  {title} {\bibinfo {title} {Elastic strips and
			differntial geometry},\ }\href
	{https://citeseerx.ist.psu.edu/viewdoc/download?doi=10.1.1.109.900&rep=rep1&type=pdf}
	{\bibfield  {journal} {\bibinfo  {journal} {Rend. Sem. Mat. Univ. Pol.
				Torino}\ }\textbf {\bibinfo {volume} {63}},\ \bibinfo {pages} {179} (\bibinfo
		{year} {2005})}\BibitemShut {NoStop}%
	\bibitem [{\citenamefont {Sabitov}(2007)}]{sabitov:isometric}%
	\BibitemOpen
	\bibfield  {author} {\bibinfo {author} {\bibfnamefont {I.~K.}\ \bibnamefont
			{Sabitov}},\ }\bibfield  {title} {\bibinfo {title} {Isometric immersions and
			embeddings of a flat {M}{\"o}bius strip in {E}uclidean spaces},\ }\href
	{https://doi.org/10.1070/im2007v071n05abeh002376} {\bibfield  {journal}
		{\bibinfo  {journal} {Izvestiya: Mathematics}\ }\textbf {\bibinfo {volume}
			{71}},\ \bibinfo {pages} {1049} (\bibinfo {year} {2007})}\BibitemShut
	{NoStop}%
	\bibitem [{\citenamefont {Starostin}\ and\ \citenamefont {van~der
			Heijden}(2007{\natexlab{a}})}]{starostin:shape}%
	\BibitemOpen
	\bibfield  {author} {\bibinfo {author} {\bibfnamefont {E.}~\bibnamefont
			{Starostin}}\ and\ \bibinfo {author} {\bibfnamefont {G.~H.~M.}\ \bibnamefont
			{van~der Heijden}},\ }\bibfield  {title} {\bibinfo {title} {The shape of a
			{M}{\"o}bius strip},\ }\href
	{https://doi.org/https://doi.org/10.1038/nmat1929} {\bibfield  {journal}
		{\bibinfo  {journal} {Nature Mater.}\ }\textbf {\bibinfo {volume} {6}},\
		\bibinfo {pages} {563} (\bibinfo {year} {2007}{\natexlab{a}})}\BibitemShut
	{NoStop}%
	\bibitem [{\citenamefont {Starostin}\ and\ \citenamefont {van~der
			Heijden}(2007{\natexlab{b}})}]{starostin:equilibrium_2007}%
	\BibitemOpen
	\bibfield  {author} {\bibinfo {author} {\bibfnamefont {E.~L.}\ \bibnamefont
			{Starostin}}\ and\ \bibinfo {author} {\bibfnamefont {G.~H.~M.}\ \bibnamefont
			{van~der Heijden}},\ }\bibfield  {title} {\bibinfo {title} {The equilibrium
			shape of an elastic developable {M}{\"o}bius strip},\ }\href
	{https://doi.org/https://doi.org/10.1002/pamm.200700858} {\bibfield
		{journal} {\bibinfo  {journal} {Proc. Appl. Math. Mech.}\ }\textbf {\bibinfo
			{volume} {7}},\ \bibinfo {pages} {2020115} (\bibinfo {year}
		{2007}{\natexlab{b}})}\BibitemShut {NoStop}%
	\bibitem [{\citenamefont {Kurono}\ and\ \citenamefont
		{Umehara}(2008)}]{kurono:flat}%
	\BibitemOpen
	\bibfield  {author} {\bibinfo {author} {\bibfnamefont {Y.}~\bibnamefont
			{Kurono}}\ and\ \bibinfo {author} {\bibfnamefont {M.}~\bibnamefont
			{Umehara}},\ }\bibfield  {title} {\bibinfo {title} {Flat {M}{\"o}bius strips
			of given isotopy type in ${R}^3$ whose centerlines are geodesics or lines of
			curvature},\ }\href
	{https://doi.org/https://doi.org/10.1007/s10711-008-9248-y} {\bibfield
		{journal} {\bibinfo  {journal} {Geom, Dedicata}\ }\textbf {\bibinfo {volume}
			{134}},\ \bibinfo {pages} {109} (\bibinfo {year} {2008})}\BibitemShut
	{NoStop}%
	\bibitem [{\citenamefont {Chubelaschwili}\ and\ \citenamefont
		{Pinkall}(2010)}]{chubelaschwili:elastic}%
	\BibitemOpen
	\bibfield  {author} {\bibinfo {author} {\bibfnamefont {D.}~\bibnamefont
			{Chubelaschwili}}\ and\ \bibinfo {author} {\bibfnamefont {U.}~\bibnamefont
			{Pinkall}},\ }\bibfield  {title} {\bibinfo {title} {Elastic strips},\ }\href
	{https://doi.org/https://doi.org/10.1007/s00229-010-0369-x} {\bibfield
		{journal} {\bibinfo  {journal} {Manuscripta Math.}\ }\textbf {\bibinfo
			{volume} {133}},\ \bibinfo {pages} {307} (\bibinfo {year}
		{2010})}\BibitemShut {NoStop}%
	\bibitem [{\citenamefont {Naokawa}(2013)}]{naokawa:extrinsically}%
	\BibitemOpen
	\bibfield  {author} {\bibinfo {author} {\bibfnamefont {K.}~\bibnamefont
			{Naokawa}},\ }\bibfield  {title} {\bibinfo {title} {{Extrinsically flat
				{M}{\"o}bius strips on given knots in 3-dimensional spaceform}},\ }\href
	{https://doi.org/10.2748/tmj/1378991020} {\bibfield  {journal} {\bibinfo
			{journal} {Tohoku Math. J.}\ }\textbf {\bibinfo {volume} {65}},\ \bibinfo
		{pages} {341} (\bibinfo {year} {2013})}\BibitemShut {NoStop}%
	\bibitem [{\citenamefont {Dias}\ and\ \citenamefont
		{Audoly}(2014)}]{dias:non-linear}%
	\BibitemOpen
	\bibfield  {author} {\bibinfo {author} {\bibfnamefont {M.~A.}\ \bibnamefont
			{Dias}}\ and\ \bibinfo {author} {\bibfnamefont {B.}~\bibnamefont {Audoly}},\
	}\bibfield  {title} {\bibinfo {title} {A non-linear rod model for folded
			elastic strips},\ }\href
	{https://doi.org/https://doi.org/10.1016/j.jmps.2013.08.012} {\bibfield
		{journal} {\bibinfo  {journal} {J. Mech. Phys. Solids}\ }\textbf {\bibinfo
			{volume} {62}},\ \bibinfo {pages} {57} (\bibinfo {year} {2014})}\BibitemShut
	{NoStop}%
	\bibitem [{\citenamefont {Kirby}\ and\ \citenamefont
		{Fried}(2015)}]{kirby:gamma-limit}%
	\BibitemOpen
	\bibfield  {author} {\bibinfo {author} {\bibfnamefont {N.~O.}\ \bibnamefont
			{Kirby}}\ and\ \bibinfo {author} {\bibfnamefont {E.}~\bibnamefont {Fried}},\
	}\bibfield  {title} {\bibinfo {title} {Gamma-limit of a model for the elastic
			energy of an inextensible ribbon},\ }\href
	{https://doi.org/https://doi.org/10.1007/s10659-014-9475-4} {\bibfield
		{journal} {\bibinfo  {journal} {J. Elast}\ }\textbf {\bibinfo {volume}
			{119}},\ \bibinfo {pages} {35} (\bibinfo {year} {2015})}\BibitemShut
	{NoStop}%
	\bibitem [{\citenamefont {Starostin}\ and\ \citenamefont {van~der
			Heijden}(2015)}]{starostin:equilibrium_2015}%
	\BibitemOpen
	\bibfield  {author} {\bibinfo {author} {\bibfnamefont {E.~L.}\ \bibnamefont
			{Starostin}}\ and\ \bibinfo {author} {\bibfnamefont {G.~H.~M.}\ \bibnamefont
			{van~der Heijden}},\ }\bibfield  {title} {\bibinfo {title} {Equilibrium
			shapes with stress localisation for inextensible elastic {M}{\"o}bius and
			other strips.},\ }\href
	{https://doi.org/https://doi.org/10.1007/s10659-014-9495-0} {\bibfield
		{journal} {\bibinfo  {journal} {J. Elast}\ }\textbf {\bibinfo {volume}
			{119}},\ \bibinfo {pages} {67} (\bibinfo {year} {2015})}\BibitemShut
	{NoStop}%
	\bibitem [{\citenamefont {Shen}\ \emph {et~al.}(2015)\citenamefont {Shen},
		\citenamefont {Huang}, \citenamefont {Chen},\ and\ \citenamefont
		{Bao}}]{shen:geometrically}%
	\BibitemOpen
	\bibfield  {author} {\bibinfo {author} {\bibfnamefont {Z.}~\bibnamefont
			{Shen}}, \bibinfo {author} {\bibfnamefont {J.}~\bibnamefont {Huang}},
		\bibinfo {author} {\bibfnamefont {W.}~\bibnamefont {Chen}},\ and\ \bibinfo
		{author} {\bibfnamefont {H.}~\bibnamefont {Bao}},\ }\bibfield  {title}
	{\bibinfo {title} {Geometrically exact simulation of inextensible ribbon},\
	}\href {https://doi.org/https://doi.org/10.1111/cgf.12753} {\bibfield
		{journal} {\bibinfo  {journal} {Comp. Graph. Forum}\ }\textbf {\bibinfo
			{volume} {34}},\ \bibinfo {pages} {145} (\bibinfo {year} {2015})}\BibitemShut
	{NoStop}%
	\bibitem [{\citenamefont {Scholtes}\ \emph {et~al.}(2019)\citenamefont
		{Scholtes}, \citenamefont {Schumacher},\ and\ \citenamefont
		{Wardetzky}}]{scholtes:variational}%
	\BibitemOpen
	\bibfield  {author} {\bibinfo {author} {\bibfnamefont {S.}~\bibnamefont
			{Scholtes}}, \bibinfo {author} {\bibfnamefont {H.}~\bibnamefont
			{Schumacher}},\ and\ \bibinfo {author} {\bibfnamefont {M.}~\bibnamefont
			{Wardetzky}},\ }\href@noop {} {\bibinfo {title} {Variational convergence of
			discrete elasticae}} (\bibinfo {year} {2019}),\ \Eprint
	{https://arxiv.org/abs/1901.02228} {arXiv:1901.02228 [math.NA]} \BibitemShut
	{NoStop}%
	\bibitem [{\citenamefont {Wunderlich}(1962)}]{wunderlich:abwickelbares}%
	\BibitemOpen
	\bibfield  {author} {\bibinfo {author} {\bibfnamefont {W.}~\bibnamefont
			{Wunderlich}},\ }\bibfield  {title} {\bibinfo {title} {\"{U}ber ein
			abwickelbares {M}\"obiusband},\ }\href
	{https://doi.org/https://doi.org/10.1007/BF01299052} {\bibfield  {journal}
		{\bibinfo  {journal} {Monatsh. fur Math}\ }\textbf {\bibinfo {volume} {66}},\
		\bibinfo {pages} {276} (\bibinfo {year} {1962})},\ \bibinfo {note} {see
		\cite{todres:translation} for an {E}nglish translation.}\BibitemShut {Stop}%
	\bibitem [{\citenamefont
		{Sadowsky}(1930{\natexlab{a}})}]{sadowsky:elementarer}%
	\BibitemOpen
	\bibfield  {author} {\bibinfo {author} {\bibfnamefont {M.}~\bibnamefont
			{Sadowsky}},\ }\bibfield  {title} {\bibinfo {title} {Ein elementarer {B}eweis
			f\"ur die {E}xistenz eines abwickelbaren \textsc{M\"obius}schen {B}andes und
			die {Z}ur\"uckf\"uhrung des geometrischen {P}roblems auf ein
			{V}ariationsproblem},\ }\href@noop {} {\bibfield  {journal} {\bibinfo
			{journal} {Sitzungsberichte der {P}reussischen {A}kademie der
				{W}issenschaften, physikalisch-mathematische {K}lasse}\ }\textbf {\bibinfo
			{volume} {22}},\ \bibinfo {pages} {412} (\bibinfo {year}
		{1930}{\natexlab{a}})},\ \bibinfo {note} {see \cite{hinz:translation} for an
		{E}nglish translation.}\BibitemShut {Stop}%
	\bibitem [{\citenamefont {Sadowsky}(1929)}]{sadowsky:differentialgleichungen}%
	\BibitemOpen
	\bibfield  {author} {\bibinfo {author} {\bibfnamefont {M.}~\bibnamefont
			{Sadowsky}},\ }\bibfield  {title} {\bibinfo {title} {Die
			differentialgleichungen des \textsc{M\"obius}schen bandes},\ }\href@noop {}
	{\bibfield  {journal} {\bibinfo  {journal} {Jahresbericht der {D}eutschen
				{M}athermatiker-{V}ereinigung, 2. {A}bt. {H}eft 5/8, {J}ahresversammlung vom
				16. bis 23. {S}eptember}\ }\textbf {\bibinfo {volume} {39}},\ \bibinfo
		{pages} {49} (\bibinfo {year} {1929})},\ \bibinfo {note} {see
		\cite{hinz:translation_bis} for an {E}nglish translation.}\BibitemShut
	{Stop}%
	\bibitem [{\citenamefont {Sadowsky}(1930{\natexlab{b}})}]{sadowsky:theorie}%
	\BibitemOpen
	\bibfield  {author} {\bibinfo {author} {\bibfnamefont {M.}~\bibnamefont
			{Sadowsky}},\ }\bibfield  {title} {\bibinfo {title} {Theorie der elastisch
			biegsamen undehnbaren {B}\"ander mit {A}nwendungen auf das
			\textsc{M\"obius}’sche {B}and},\ }in\ \href@noop {} {\emph {\bibinfo
			{booktitle} {Proc. of the 3rd {I}nt. {C}ongress of {A}pplied {M}echanics}}},\
	Vol.~\bibinfo {volume} {2},\ \bibinfo {editor} {edited by\ \bibinfo {editor}
		{\bibfnamefont {A.~C.~W.}\ \bibnamefont {Oseen}}\ and\ \bibinfo {editor}
		{\bibfnamefont {W.}~\bibnamefont {Weibull}}}\ (\bibinfo  {publisher} {{AB}.
		{S}veriges {L}itografiska {T}ryckerier},\ \bibinfo {address} {Stockholm,
		Sweden},\ \bibinfo {year} {1930})\ pp.\ \bibinfo {pages} {444--451},\
	\bibinfo {note} {see \cite{hinz:translation_ter} for an {E}nglish
		translation.}\BibitemShut {Stop}%
	\bibitem [{\citenamefont {Chen}\ \emph
		{et~al.}(2018{\natexlab{b}})\citenamefont {Chen}, \citenamefont {Fosdick},\
		and\ \citenamefont {Fried}}]{chen:representation}%
	\BibitemOpen
	\bibfield  {author} {\bibinfo {author} {\bibfnamefont {Y.-C.}\ \bibnamefont
			{Chen}}, \bibinfo {author} {\bibfnamefont {R.}~\bibnamefont {Fosdick}},\ and\
		\bibinfo {author} {\bibfnamefont {E.}~\bibnamefont {Fried}},\ }\bibfield
	{title} {\bibinfo {title} {Representation of a smooth isometric deformation
			of a planar material region into a curved surface},\ }\href
	{https://doi.org/https://doi.org/10.1007/s10659-017-9637-2} {\bibfield
		{journal} {\bibinfo  {journal} {J. Elast.}\ }\textbf {\bibinfo {volume}
			{130}},\ \bibinfo {pages} {145} (\bibinfo {year}
		{2018}{\natexlab{b}})}\BibitemShut {NoStop}%
	\bibitem [{\citenamefont {Freddi}\ \emph {et~al.}(2016)\citenamefont {Freddi},
		\citenamefont {Hornung}, \citenamefont {Mora},\ and\ \citenamefont
		{Paroni}}]{freddi:corrected}%
	\BibitemOpen
	\bibfield  {author} {\bibinfo {author} {\bibfnamefont {L.}~\bibnamefont
			{Freddi}}, \bibinfo {author} {\bibfnamefont {P.}~\bibnamefont {Hornung}},
		\bibinfo {author} {\bibfnamefont {M.~G.}\ \bibnamefont {Mora}},\ and\
		\bibinfo {author} {\bibfnamefont {R.}~\bibnamefont {Paroni}},\ }\bibfield
	{title} {\bibinfo {title} {A corrected {S}adowsky functional for inextensible
			elastic ribbons},\ }\href
	{https://doi.org/https://doi.org/10.1007/s10659-015-9551-4} {\bibfield
		{journal} {\bibinfo  {journal} {J. Elast.}\ }\textbf {\bibinfo {volume}
			{123}},\ \bibinfo {pages} {125} (\bibinfo {year} {2016})}\BibitemShut
	{NoStop}%
	\bibitem [{\citenamefont {Goriely}\ \emph {et~al.}(2022)\citenamefont
		{Goriely}, \citenamefont {Moulton},\ and\ \citenamefont
		{Mihai}}]{goriely:rod}%
	\BibitemOpen
	\bibfield  {author} {\bibinfo {author} {\bibfnamefont {A.}~\bibnamefont
			{Goriely}}, \bibinfo {author} {\bibfnamefont {D.~E.}\ \bibnamefont
			{Moulton}},\ and\ \bibinfo {author} {\bibfnamefont {L.~A.}\ \bibnamefont
			{Mihai}},\ }\bibfield  {title} {\bibinfo {title} {A rod theory for liquid
			crystalline elastomers},\ }\href
	{https://doi.org/https://doi.org/10.1007/s10659-021-09875-z} {\bibfield
		{journal} {\bibinfo  {journal} {J. Elast.}\ } (\bibinfo {year}
		{2022})}\BibitemShut {NoStop}%
	\bibitem [{\citenamefont {Hornung}(2010)}]{hornung:euler}%
	\BibitemOpen
	\bibfield  {author} {\bibinfo {author} {\bibfnamefont {P.}~\bibnamefont
			{Hornung}},\ }\bibfield  {title} {\bibinfo {title} {Euler-{L}agrange
			equations for variational problems on space curves},\ }\href
	{https://doi.org/10.1103/PhysRevE.81.066603} {\bibfield  {journal} {\bibinfo
			{journal} {Phys. Rev. E}\ }\textbf {\bibinfo {volume} {81}},\ \bibinfo
		{pages} {066603} (\bibinfo {year} {2010})}\BibitemShut {NoStop}%
	\bibitem [{\citenamefont {Antman}(1995)}]{antman:nonlinear}%
	\BibitemOpen
	\bibfield  {author} {\bibinfo {author} {\bibfnamefont {S.~S.}\ \bibnamefont
			{Antman}},\ }\href@noop {} {\emph {\bibinfo {title} {Nonlinear Problems of
				Elasticity}}},\ \bibinfo {series} {Applied Mathematical Sciences}, Vol.\
	\bibinfo {volume} {107}\ (\bibinfo  {publisher} {Springer},\ \bibinfo
	{address} {New York},\ \bibinfo {year} {1995})\BibitemShut {NoStop}%
	\bibitem [{\citenamefont {Beer}(1852)}]{beer:bestimmung}%
	\BibitemOpen
	\bibfield  {author} {\bibinfo {author} {\bibfnamefont {A.}~\bibnamefont
			{Beer}},\ }\bibfield  {title} {\bibinfo {title} {Bestimmung der {A}bsorption
			des rothen {L}ichts in farbigen {F}l\"ussigkeiten},\ }\href
	{https://doi.org/https://doi.org/10.1002/andp.18521620505} {\bibfield
		{journal} {\bibinfo  {journal} {Ann. Phys.}\ }\textbf {\bibinfo {volume}
			{162}},\ \bibinfo {pages} {78} (\bibinfo {year} {1852})}\BibitemShut
	{NoStop}%
	\bibitem [{\citenamefont {Fox}(2010)}]{fox:otptical}%
	\BibitemOpen
	\bibfield  {author} {\bibinfo {author} {\bibfnamefont {M.}~\bibnamefont
			{Fox}},\ }\href@noop {} {\emph {\bibinfo {title} {Optical Properties of
				Solids}}},\ \bibinfo {edition} {2nd}\ ed.\ (\bibinfo  {publisher} {Oxford
		University Press},\ \bibinfo {address} {Oxford},\ \bibinfo {year}
	{2010})\BibitemShut {NoStop}%
	\bibitem [{\citenamefont {Camacho-Lopez}\ \emph {et~al.}(2004)\citenamefont
		{Camacho-Lopez}, \citenamefont {Finkelmann}, \citenamefont {Palffy-Muhoray},\
		and\ \citenamefont {Shelley}}]{camacho-lopez:fast}%
	\BibitemOpen
	\bibfield  {author} {\bibinfo {author} {\bibfnamefont {M.}~\bibnamefont
			{Camacho-Lopez}}, \bibinfo {author} {\bibfnamefont {H.}~\bibnamefont
			{Finkelmann}}, \bibinfo {author} {\bibfnamefont {P.}~\bibnamefont
			{Palffy-Muhoray}},\ and\ \bibinfo {author} {\bibfnamefont {M.}~\bibnamefont
			{Shelley}},\ }\bibfield  {title} {\bibinfo {title} {Fast liquid-crystal
			elastomer swims into the dark},\ }\href
	{https://doi.org/https://doi.org/10.1038/nmat1118} {\bibfield  {journal}
		{\bibinfo  {journal} {Nature Mater.}\ }\textbf {\bibinfo {volume} {3}},\
		\bibinfo {pages} {307} (\bibinfo {year} {2004})}\BibitemShut {NoStop}%
	\bibitem [{\citenamefont {Liu}\ \emph {et~al.}(2020)\citenamefont {Liu},
		\citenamefont {del Pozo}, \citenamefont {Mohseninejad}, \citenamefont
		{Debije}, \citenamefont {Broer},\ and\ \citenamefont
		{Schenning}}]{liu:light}%
	\BibitemOpen
	\bibfield  {author} {\bibinfo {author} {\bibfnamefont {L.}~\bibnamefont
			{Liu}}, \bibinfo {author} {\bibfnamefont {M.}~\bibnamefont {del Pozo}},
		\bibinfo {author} {\bibfnamefont {F.}~\bibnamefont {Mohseninejad}}, \bibinfo
		{author} {\bibfnamefont {M.~G.}\ \bibnamefont {Debije}}, \bibinfo {author}
		{\bibfnamefont {D.~J.}\ \bibnamefont {Broer}},\ and\ \bibinfo {author}
		{\bibfnamefont {A.~P. H.~J.}\ \bibnamefont {Schenning}},\ }\bibfield  {title}
	{\bibinfo {title} {Light tracking and light guiding fiber arrays by adjusting
			the location of photoresponsive azobenzene in liquid crystal networks},\
	}\href {https://doi.org/https://doi.org/10.1002/adom.202000732} {\bibfield
		{journal} {\bibinfo  {journal} {Adv. Optical Mater.}\ }\textbf {\bibinfo
			{volume} {8}},\ \bibinfo {pages} {2000732} (\bibinfo {year}
		{2020})}\BibitemShut {NoStop}%
	\bibitem [{\citenamefont {Sluckin}\ \emph {et~al.}(2004)\citenamefont
		{Sluckin}, \citenamefont {Dunmur},\ and\ \citenamefont
		{Stegemeyer}}]{sluckin:crystal}%
	\BibitemOpen
	\bibfield  {author} {\bibinfo {author} {\bibfnamefont {T.~J.}\ \bibnamefont
			{Sluckin}}, \bibinfo {author} {\bibfnamefont {D.~A.}\ \bibnamefont
			{Dunmur}},\ and\ \bibinfo {author} {\bibfnamefont {H.}~\bibnamefont
			{Stegemeyer}},\ }\href@noop {} {\emph {\bibinfo {title} {Crystals that
				Flow}}}\ (\bibinfo  {publisher} {Taylor \& Francis},\ \bibinfo {address}
	{London, New York},\ \bibinfo {year} {2004})\BibitemShut {NoStop}%
	\bibitem [{\citenamefont {R.Todres}(2015)}]{todres:translation}%
	\BibitemOpen
	\bibfield  {author} {\bibinfo {author} {\bibnamefont {R.Todres}},\ }\bibfield
	{title} {\bibinfo {title} {Translation of {W}. {W}underlich’s ‘{O}n a
			developable {M}{\"o}bius band’},\ }\href
	{https://doi.org/doi:10.1007/s10659-014-9489-y} {\bibfield  {journal}
		{\bibinfo  {journal} {J. Elast.}\ }\textbf {\bibinfo {volume} {119}},\
		\bibinfo {pages} {23} (\bibinfo {year} {2015})}\BibitemShut {NoStop}%
	\bibitem [{\citenamefont {Hinz}\ and\ \citenamefont
		{Fried}(2015{\natexlab{a}})}]{hinz:translation}%
	\BibitemOpen
	\bibfield  {author} {\bibinfo {author} {\bibfnamefont {D.~F.}\ \bibnamefont
			{Hinz}}\ and\ \bibinfo {author} {\bibfnamefont {E.}~\bibnamefont {Fried}},\
	}\bibfield  {title} {\bibinfo {title} {Translation of {M}ichael
			{S}adowsky’s paper ‘{A}n elementary proof for the existence of a
			developable \textsc{M\"obius} band and the attribution of the geometric
			problem to a variational problem’},\ }\href
	{https://doi.org/doi:10.1007/s10659-014-9490-5} {\bibfield  {journal}
		{\bibinfo  {journal} {J. Elast.}\ }\textbf {\bibinfo {volume} {119}},\
		\bibinfo {pages} {3} (\bibinfo {year} {2015}{\natexlab{a}})}\BibitemShut
	{NoStop}%
	\bibitem [{\citenamefont {Hinz}\ and\ \citenamefont
		{Fried}(2015{\natexlab{b}})}]{hinz:translation_bis}%
	\BibitemOpen
	\bibfield  {author} {\bibinfo {author} {\bibfnamefont {D.~F.}\ \bibnamefont
			{Hinz}}\ and\ \bibinfo {author} {\bibfnamefont {E.}~\bibnamefont {Fried}},\
	}\bibfield  {title} {\bibinfo {title} {Translation of {M}ichael
			{S}adowsky’s paper ‘{T}he differential equations of the \textsc{M\"obius}
			band’},\ }\href {https://doi.org/doi:10.1007/s10659-014-9491-4} {\bibfield
		{journal} {\bibinfo  {journal} {J. Elast.}\ }\textbf {\bibinfo {volume}
			{119}},\ \bibinfo {pages} {19} (\bibinfo {year}
		{2015}{\natexlab{b}})}\BibitemShut {NoStop}%
	\bibitem [{\citenamefont {Hinz}\ and\ \citenamefont
		{Fried}(2015{\natexlab{c}})}]{hinz:translation_ter}%
	\BibitemOpen
	\bibfield  {author} {\bibinfo {author} {\bibfnamefont {D.~F.}\ \bibnamefont
			{Hinz}}\ and\ \bibinfo {author} {\bibfnamefont {E.}~\bibnamefont {Fried}},\
	}\bibfield  {title} {\bibinfo {title} {Translation of {M}ichael
			{S}adowsky’s paper ‘{T}heory of elastically bendable inextensible bands
			with applications to the \textsc{M\"obius} band’},\ }\href
	{https://doi.org/doi:10.1007/s10659-014-9492-3} {\bibfield  {journal}
		{\bibinfo  {journal} {J. Elast.}\ }\textbf {\bibinfo {volume} {119}},\
		\bibinfo {pages} {7} (\bibinfo {year} {2015}{\natexlab{c}})}\BibitemShut
	{NoStop}%
\end{thebibliography}

%

\end{document}